\documentclass[fleqn,usenatbib]{mnras}

\usepackage{newtxtext,newtxmath}

\usepackage[T1]{fontenc}
\usepackage{ae,aecompl}

\usepackage{graphicx}	

\newcommand{\Msun}{\rm M_{\sun}}

\newcommand{\Gyr}{\rm Gyr}

\newcommand{\ergs}{\mbox{erg}~\mbox{s}^{-1}}
\newcommand{\kpc}{\rm kpc}
\newcommand{\pc}{\rm pc}
\newcommand{\ckpc}{\rm ckpc}

\newcommand{\cMpc}{\rm cMpc}
\newcommand{\Mpc}{\rm Mpc}

\newcommand{\lsim}{\mathrel{\hbox{\rlap{\lower.55ex\hbox{$\sim$}} \kern-.3em\raise.4ex\hbox{$<$}}}}
\newcommand{\gsim}{\mathrel{\hbox{\rlap{\lower.55ex\hbox{$\sim$}} \kern-.3em\raise.4ex\hbox{$>$}}}}

\newcommand{\Ato}{A_{2} }
\newcommand{\Atot}{A_{\rm 2, tot} }
\newcommand{\Amax}{A_{\rm 2,max} }

\newcommand{\rmax}{r_{\rm max} }
\newcommand{\rbar}{r_{\rm bar} }
\newcommand{\rhalf}{r_{\rm 50,*} }
\newcommand{\tnorm}{t_{\rm norm} }
\newcommand{\tbar}{t_{\rm bar} }
\newcommand{\TNG}{TNG100}
\newcommand{\TNGF}{TNG50}
\newcommand{\SFE}{\rm SFE}
\newcommand{\SFEdr}{\rm SFE(\rm dr)/SFE_{\rm gal} }

\title{The buildup of strongly barred galaxies in the \TNG~ simulation}

\author[Rosas-Guevara Y. M. et al.]{Yetli Rosas-Guevara,$^{1}$\thanks{E-mail: yetli.rosas@dipc.org}
Silvia Bonoli,$^{1,2}$
Massimo Dotti, $^{3,4}$
Tommaso Zana,$^{5,6}$ 
 \newauthor Dylan Nelson,$^{7}$
Annalisa Pillepich,$^8$
Luis C. Ho,$^{9,10}$
David Izquierdo-Villalba,$^{11}$
\newauthor Lars Hernquist$^{12}$
and R\"{u}ediger Pakmor$^{6}$ \\
$^{1}$ Donostia International Physics Centre (DIPC), Paseo Manuel de Lardizabal 4, 20018 Donostia-San Sebastian, Spain\\
$^{2}$IKERBASQUE, Basque Foundation for Science, E-48013, Bilbao, Spain\\
$^{3}$Dipartimento di Fisica G. Occhialini, Universit\`{a} di Milano-Bicocca, Piazza della Scienza 3, IT-20126 Milano, Italy\\
$^{4}$INFN, Sezione di Milano-Bicocca, Piazza della Scienza 3, IT-20126 Milano, Italy\\
$^5$DiSAT, Universit\`{a} degli Studi dell'Insubria, Via Valleggio 11, IT-22100 Como, Italy\\
$^6$Scuola Normale Superiore, Piazza dei Cavalieri 7, IT-56126 Pisa, Italy \\
$^7$ Max Planck Institute for Astrophysics, Karl-Schwarzschild-Str. 1, 85741 Garching bei Munchen, Germany \\
$^8$Max-Planck-Institut f\"{u}r Astronomie, K\"{o}nigstuhl 17, D-69117 Heidelberg, Germany \\
$^9$Kavli Institute for Astronomy and Astrophysics, Peking University, Beijing 100871, China \\
$^{10}$Department of Astronomy, Peking University, Beijing 100871, China \\
$^{11}$Centro de Estudios de F\'{\i}sica del Cosmos de Arag\'{o}n (CEFCA), Plaza San Juan 1, Planta-2, Teruel, 44001, Spain.\\
$^{12}$Harvard-Smithsonian Center for Astrophysics, 60 Garden Street, Cambridge 02138, MA, USA\\}

\date{Accepted XXX. Received YYY; in original form ZZZ}

\pubyear{2015}
 
\begin{document}
\label{firstpage}
\pagerange{\pageref{firstpage}--\pageref{lastpage}}
\maketitle

\begin{abstract}
We analyse the properties of strongly barred disc galaxies using the TNG100 simulation, a cosmological hydrodynamical realisation of the IllustrisTNG suite. We identify  270 disc galaxies at $z=0$ in the stellar mass range  $M_{*}=10^{10.4-11}\Msun$, of which 40 per cent are barred. Of the detected bars, more than half are strong. We find that the fraction of barred galaxies increases with stellar mass, in agreement with observational results. Strongly barred galaxies exhibit, overall, lower gas-to-stellar mass ratios compared to unbarred galaxies. The majority of barred galaxies are quenched (sSFR $\sim10^{-11.7} $yr$^{-1}$),  whereas unbarred galaxies continue to be active (sSFR $\sim10^{-10.3}$yr$^{-1}$) on the main sequence of star-forming galaxies.  We explore the evolution of strongly barred and unbarred galaxies to investigate their formation and quenching histories. We find that strong bars form between $0.5< z< 1.5$, with more massive galaxies hosting older bars.  Strong bars form in galaxies with an early-established prominent disc component, undergoing periods of enhanced star formation and black hole accretion, possibly assisted by cosmological inflows. Unbarred galaxies, on the other hand, assemble most of their mass and disc component at late times. The nuclear region of strongly barred galaxies quenches shortly after bar formation, while unbarred galaxies remain active across time.  Our findings are indicative of bar quenching, possibly assisted by nuclear feedback processes. 
 We conclude that the cosmological environment, together with small scale feedback processes, determine the chances of a galaxy to form a bar and  to rapidly quench its central region.
\end{abstract}




\section{Introduction}
Stellar bars are a common feature in the inner parts of the disc galaxies. They have been observed in more than  30 per cent  of massive disc galaxies ($ M_{*} >10^{10} M_{\odot}$) in the local Universe (e.g., \citealt{sellwood1993,masters2011}). It is believed that bars could play a crucial role in the secular evolution of disc galaxies (e.g., \citealt{debattista2004, athanassoula2005}) and also in the dynamical redistribution of gas (e.g., \citealt{athanassoula1992, romerogomez2007,berentzen2007}).  
 However, the extent to which bars are involved in regulating star formation and in which stage of galaxy evolution they become an important mechanism in shaping galaxies is still unclear. The real answer to this  might be complex  since the dynamics of the gas is highly sensitive to the local environment of the interstellar medium and to the conditions under which a bar forms and grows (e.g., \citealt{englmaier2000,fragkoudi2016}).
 
One powerful way to investigate bar-driven secular evolution in a cosmological context is to determine the fraction of barred galaxies as a function of various galaxy properties, such as stellar mass or star formation rate.  Observational studies have shown that bars are more frequently found in massive red galaxies (e.g., \citealt{barazza2008,masters2012,gavazzi2015}). In terms of their evolution, the fraction of disc galaxies with bars declines rapidly and monotonically with increasing redshift from $\sim 0.65$  at $z\sim 0.2$ to $\sim 0.20$ at $z\sim 0.8$ \citep{sheth2008}. 
For instance, \cite*{sheth2012}, 
analysing a sample of disc galaxies in the COSMOS survey, suggest that the steep decline seen could be due to an evolution in the dynamics of the discs: galaxies with a stellar bar are likely to reside in a massive and cold disc  whereas galaxies with a (dynamically) hot disc do not develop bars. Besides, hot discs are more common at higher redshifts \citep[e.g.,][]{law2012,kraljic2012,pillepich2019}.  The authors also note that not all cold and massive disc galaxies have bars, suggesting that a secondary process, such as the interaction history of the galaxy, might be relevant in determining which galaxies develop bars. Although there is a consensus in the community about the evolution of barred galaxy fraction,  \cite{erwin2018} points out that the observed bar fractions in the local universe could be underestimated, especially in low mass galaxies.  The author  also remarks that high redshift bar fractions could be underestimated due to a combination of poor angular resolution and the correlation between bar size and stellar mass.

 Another feature seen in observations is the possible connection between the star formation of a galaxy  and the presence of a bar (e.g., \citealt{laurikainen2004,jogee2005,masters2010,james2016}).
 Using the data set from GalaxyZoo-2, \cite{cheung2013} find that the probability of a galaxy hosting a bar is anticorrelated with the specific star formation rate regardless of the stellar mass or bulge prominence. Similarly,
 \cite{gavazzi2015} identify a threshold in the stellar mass above which a sharp increase in the fraction of visually classified strong bars and a concomitant decrease in the specific star formation rate (sSFR) is observed in the local Universe\footnote{The increase in the strongly barred galaxy fraction with mass has been confirmed through an automated analysis by \cite{consolandi16}.}. 
 Such a threshold in the sSFR has been found to increase with redshift\footnote{The limited resolution of the data prevented \cite{gavazzi2015} from checking if the same evolution in the mass threshold also applies to the fraction of barred galaxies.}. \cite*{gavazzi2015} suggest that strong bars may be responsible for the \textit{quenching}  in star formation observed at high redshift. Strong bars can induce inflows into the central part of the galaxy producing a starburst or feeding the central black hole. As a result, a decline in the central star formation rate takes place. This mechanism is referred to as \textit{bar-driven quenching}. Similar conclusions are reached by \cite*{kim2017} who study strongly barred galaxies in the SDSS and find their star formation activity, on average, is lower than that of unbarred galaxies. 
However, the long-term relevance of bar-driven quenching is still unclear since bars could be dissolved through time. Besides, there are other mechanisms, such as feedback, that can quench star formation in galaxies \citep{george2019}. 

 Idealised simulations of disc galaxies have shed light on the possible mechanisms that could affect the formation and evolution of bars. \cite{ostriker1973}  observe that the presence of a dominant spherical component is needed to prevent the formation of bars in discs.
 \citealt{athanassoula2002, athanassoula2003, martinez2006}  have suggested different mechanisms that could affect the formation and evolution of a bar in a well-defined disc galaxy. For example, bars can cause significant exchanges of angular momentum between the stellar and gas components (e.g., \citealt{athanassoula2003}), during which the gas inside the co-rotation radius falls into the central region and prevents the inflow of gas from external regions (\citealt{athanassoula1992}). Idealised galaxy simulations including feedback, cooling and star formation \citep{athanassoula2013} have also found that bars, in the presence of large amounts of gas, form later and are weaker than in gas-poor galaxies. 

Although these studies have been very insightful, they do not take into account the effect of the large-scale environment  such as interactions with other galaxies and cosmological gas inflows.
Theoretical studies of barred galaxies in a fully cosmological context have not been  explored until recently with high-resolution zoom-in simulations of a Milky Way-type halo (e.g., \citealt{scannapieco2012,bonoli2016}).  In particular, \cite*{bonoli2016} present the zoom-in simulation \textit{ErisBH} of a Milky Way-type galaxy which is a sibling of the \textit{Eris} simulation \citep{guedes2011}, but includes black hole (BH) subgrid physics. \cite{bonoli2016} find that the simulated galaxy forms a strong bar after $z\sim1$, and they point out that  the disc in the simulation is more prone to instabilities compared to the original \textit{Eris}, possibly because of early AGN feedback which prevents the bulge from growing.    \cite*{spinoso2017}  extend this analysis in \textit{ErisBH} and find that at early stages of bar formation, the bar produces a strong torque on the gas, driving gas inflows towards the central parts (at parsec scales), briefly enhancing  star formation. They also find that the gas can be removed rapidly by the bar in the inner region, preventing any further strong star formation (see \citealt{zana2018a,zana2018c,donohoe2019} for similar analyses). 

Currently, it is possible to study the physics behind driving the evolution of a strongly barred galaxy population thanks to the new generation of cosmological hydrodynamic simulations \citep{vogelsberger2014a,schaye2015}. 
These new hydrodynamics simulations are able to reproduce many observables of a galaxy population at low redshift. In particular, there are some studies focused on  the bar population. For instance, \cite{peschken2019}, using the Illustris simulations, have suggested that a large fraction of bars ($\sim 80$ per cent) are formed by galaxy interaction events, such as mergers or flybys, and these interactions could also influence the growth of the bar under certain conditions.

Our main goal is to investigate the theoretical predictions in the build-up of strongly barred galaxies and the star formation quenching in the IllustrisTNG simulations \citep{pillepich2018b,nelson2018a,springel2018,naiman2018,marinacci2018}. In particular, we use the simulation \TNG~, which offers the best compromise between a large cosmological volume and resolution, and focus on the following questions: what are the conditions to form a strong bar? and what could possibly drive the quenching on the central region of strongly barred galaxies? By  \textit{quenching} we refer to the decline of the star formation due to any mechanism that  expels,  prevents, or rapidly consumes the gas content of a galaxy. In particular, we want to study the possibility of \textit{bar quenching}, that is the role of bar structures in modifying the star formation and gas properties of galactic discs. We base our analysis on comparing the properties between strongly barred galaxies and unbarred galaxies before and after a bar instability develops in the strongly barred galaxies. 

The paper is structured as follows. In section~\ref{sec:methodology},  we give a brief overview of the IllustrisTNG project, of our galaxy disc sample, and of our methodology to identify a bar and trace its evolution.    In section~\ref{sec:barpropz0}, we concentrate on the properties of the bar structures at $z=0$ and compare them with observations and highlight the differences between barred and unbarred galaxies in terms of star formation. In section~\ref{sec:evolution}, we track down the cosmological evolution of $z=0$ strongly barred galaxies and contrast them with the cosmological evolution of $z=0$ unbarred galaxies for a given stellar mass.  
We discuss the emerging picture of our results and discuss the limitation of our analysis in section~\ref{sec:discussion}. Finally, in section~\ref{sec:summary},  we summarise our findings.

\section{Methodology}
\label{sec:methodology}
\subsection{Overview of the simulations}
The IllustrisTNG (The Next Generation) project (\citealt{nelson2018a,naiman2018,pillepich2018b,marinacci2018,springel2018})
\footnote{\citep{nelson2019a}; http://www.tng-project.org} comprises three main cosmological, gravo-magneto-hydrodynamical simulations of galaxy formation with different volumes, ranging from $50$ to $300\,\cMpc$  with different spatial  and mass resolutions.  
The IllustrisTNG simulations are run with the moving-mesh \textsc{AREPO} code \citep{springel2010} that employs  Tree-PM approach along with a Godunov/finite volume method to discretise space.  The scheme is quasi-Langragian, second-order in both space and time. The simulations adopt the Planck cosmology parameters   with constraints  from \cite{planck2016}: $\Omega_\Lambda=0.6911$, $\Omega_{\rm m}=0.3089$, $\Omega_{\rm b}=0.0486$, $\sigma_8=0.8159$, $h=0.6774$ and $n_{s}=0.9667$  where  $\Omega_\Lambda$, $\Omega_{\rm m}$ and  $\Omega_{\rm b}$ are the average densities of matter, dark energy and baryonic matter in units  of the critical density at $z=0$,  $\sigma_8$, the square root of the linear variance,  $h$ is the Hubble parameter ($H_{o}\equiv h \,100 \rm km  s^{-1}$) and  $n_{s}$  is the scalar power-law index of the power spectrum of primordial adiabatic perturbations. The initial conditions of the simulation suite are set to $z=127$ using the Zeldovich approximation and include a uniform magnetic seed field with a comoving field strength of $10^{-14}$ Gauss.
In this paper, we focus on the simulation \TNG~ that has a comoving volume of $(110.7)^3\cMpc^3$. The setup of the simulation is provided in Table~\ref{table:simulations}.

\begin{table}
\caption{Main information of the \TNG~simulation. From top to bottom: Name of the simulation,  box side-length,  number of initial resolution elements including gas cells and  dark matter (DM) particles,  initial mass of gas cells and of  DM particles, the minimum proper softening length allowed for gas cells, the proper softening length for the collisionless particles at $z=0$, and their comoving softening length.}
\begin{center}
\begin{tabular}{|l|l|l|} 
\hline
Name   &            &   \TNG         \\     
\hline
$L$    &  [$\Mpc$]  &     $110.7$     \\

$N$    &            &   $2\times1820^3$  \\            
$m_{\rm g}$  &[$M_\odot$]&   $1.39\times 10^6$   \\        
$m_{\rm DM}$ & [$M_\odot$]&  $7.49\times10^6$  \\
$\epsilon_{\rm gas,min}$ &  $[\pc]$  &   $185$    \\
$\epsilon_{\rm DM,stars,0}$ & $[\kpc]$ &  $0.74$ \\
$\epsilon_{\rm DM,stars,z}$ & $[\ckpc]$  &  $1.48$ to $0.74$\\                    
\hline
\end{tabular}
\end{center}
\label{table:simulations}
\end{table}
\subsubsection{Subgrid physics of galaxy formation}
The subgrid physics of IllustrisTNG is partly based on its predecessor, Illustris \citep{vogelsberger2013,vogelsberger2014a,vogelsberger2014b,genel2014,nelson2015,sijacki2015}.  Significant modifications have been made to star formation feedback (winds), the growth of supermassive black holes, Active  Galactic Nuclei (AGN) Feedback, and stellar evolution and chemical enrichment. A complete description of the improvements in the subgrid physics and calibration process can be found in \cite{pillepich2018a} and \cite{weinberger2017}.   A summary of the improvements with respect to Illustris is shown in Table~1 of \citealt{pillepich2018a}. Here,  we enumerate the key physical processes  relevant to this work. 

Gas radiative mechanisms are implemented with primordial \citep{katz1996} and metal line cooling, and heating by a time-dependent ultraviolet background field from stars and luminous AGN \citep{faucher2009}. Star formation in the dense interstellar medium is treated stochastically following  an empirical Kennicutt-Schmidt relation \citep{springel2003}. Each stellar particle represents a population of stars with a common birth time following a Chabrier initial mass function.  The stellar evolution is modelled in order to calculate chemical enrichment and  mass expelled into the interstellar medium due to AGB stars, SNIa and SNII. Also, the evolution and production of ten elements (H, He, C, N, O, Ne, Mg, Si, Fe, \& Eu) are individually tracked. 

Stellar feedback is modelled by galactic-scale outflows. Wind particles are launched directly from star-forming gas with an initial wind speed that scales  with the local velocity dispersion of dark matter particles, a dependence on redshift, and limited to a minimum wind velocity value. The wind particles  are  isotropically ejected. The wind mass-loading is determined from  the available SN energy with a small  fraction  that is removed thermally.  The wind mass-loading  also depends on the metallicity of the star-forming gas cells.

Supermassive black holes are formed in massive haloes with initial black hole mass of $1.18\times10^{6}\Msun$, and can grow via two growth channels: BH mergers and gas accretion. Black holes are  Eddington-limited and allowed to accrete at the Bondi accretion rate.  

There are two modes of AGN feedback:  thermal \textit{ quasar mode}  that heats the surrounding gas of the BH at high accretion rates
\citep{springel2005, diMatteo2005}, and a kinetic \textit{wind mode} that produces a wind at low accretion rates.  The black holes are allowed to switch from \textit{quasar to wind mode} and vice versa if the Eddington ratio falls to a threshold value (set to $\lsim 0.1$) that depends on a power law of the black hole mass. The feedback energy in the \textit{ quasar mode}  is released continuously as thermal energy into the surrounding gas given by $\Delta E= 0.02 \dot{M}c^2\Delta t$. In the \textit{wind mode} case,  the energy release is kinetic and injected as a kick into the surroundings given  by  $\Delta E= \epsilon_{f,kin} \dot{M}c^2\Delta t$ where  $\epsilon_{f,kin}\lsim 0.2$, decreasing towards high densities. This allows that in the low environmental densities, AGN feedback  has the ability to weakly affect its surroundings. The model of AGN feedback described above has been shown to be responsible for the quenching of galaxies at intermediate and high mass haloes \citep{weinberger2018} and for the emergence of red passive galaxies at late times \citep{nelson2018a}. 

\subsubsection{ Galaxy Identification} 
Galaxies and their haloes  are identified as   bound substructures  using  a \textsc{FoF} and  then a \textsc{SUBFIND} algorithm \citep{springel2001} and  connected over time  by the \textsc{Sublink} merger tree algorithm \citep{rodriguezgomez2015}. 
 
\subsection{Parent disc-galaxy sample}
\begin{figure*}	
\begin{tabular}{cc}
\includegraphics[width=\columnwidth]{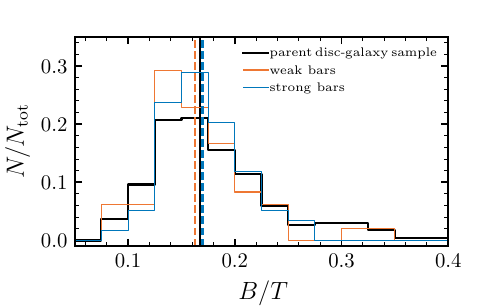} &
\includegraphics[width=\columnwidth]{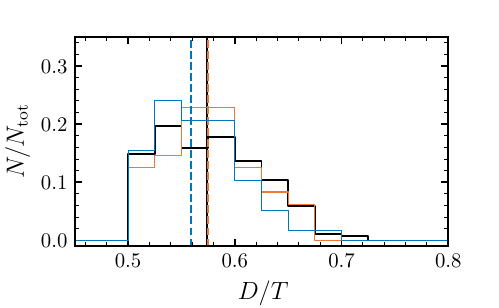}  \\
\end{tabular} 
\caption{Distribution of $B/T$ and $D/T$ for TNG100 galaxies calculated within an aperture of $10 R_{\rm 50,*}$.The black solid lines show the distribution of the parent-disc galaxies while the blue and orange solid lines correspond to the galaxies with strong and weak bars in the parent-disc galaxies, respectively.  The vertical solid lines indicate the median values in the $B/T$ and $D/T$ distributions for the parent disc galaxies, the vertical dashed lines to the medians in weak  and strong bars. The figure highlights that disc galaxies have low values of $B/T$.}
\label{fig:morphgal}
\end{figure*}

We concentrate our study on disc-dominated galaxies with  enough resolution elements (more than $10^4$ stellar particles within the half mass radius) to  study the different morphological components (such as discs, bulges and bars) and their internal structure with high accuracy.
 We start, thus imposing a low stellar mass cut of $10^{10.4}\Msun$ (within $2r_{50,*} $ where $r_{50,*}$ is the stellar half-mass radius) which results in more than 2500 galaxies in the TNG100 simulation. To identify disc galaxies,  we use the kinematic bulge-to-disc decomposition described in \cite{genel2015} (provided by the IllustrisTNG team), which follows \cite{marinacci2014} and \cite{abadi2003}. In this algorithm, for each stellar particle within $10r_{50,*}$ the circularity parameter is defined as 
\begin{equation}
\epsilon = \dfrac{J_z}{J(E)},
\label{eq:circ}
\end{equation}
where  $J_{z}$ is the specific angular momentum of the particle around the symmetry axis and  $J(E)$ is the maximum specific angular momentum possible at the specific binding energy of each stellar particle. The stellar disc mass is based on the stellar particles with $\epsilon>0.7$, while the bulge mass is defined as twice the  mass of stellar particles with a circularity parameter $\epsilon<0$. In the TNG100 simulation, we find more than 300 disc galaxies. Note that \cite{naiman2018} find a higher number of Milky Way disc galaxies ($\sim 800$) in the same simulation. This difference is because the authors impose a minimum cut in halo mass instead of stellar mass, that includes a significant number of galaxies with a stellar mass smaller than the minimum stellar cut we consider here.
We find that not all galaxies have a clear morphological classification, with a large number of stellar particles not belonging to either the bulge or disc.  Those are mainly galaxies in an unrelaxed state. Given that our goal is to study stable bar structures and their secular effects on the host galaxies, we limit our analysis to galaxies with a well-defined morphology. Therefore, we include in our parent sample only galaxies with $(D/T+ B/T)\geq0.7$  where  $D/T$ is the stellar disc-to-total mass ratio and $B/T$ is the stellar bulge-to-total mass ratio. Finally, we select as disc-dominated the galaxies with $D/T\geq0.5$. We also try other apertures to define morphology, such as considering all the stellar particles in the halo and $2\rhalf$, finding similar results. 

When all these cuts are applied, we end up with a parent sample of $270$ disc-dominated galaxies of which 213 are centrals and 57 are satellites.  Fig.~\ref{fig:morphgal} shows the distribution of B/T and D/T ratios of this parent disc-galaxy sample. By construction, the D/T ratios are always above $0.5$, with a median close to $0.6$. Bulges are relatively small, with typical $B/T$ $\sim 0.15$.  Barred galaxies (see subsection~\ref{subsec:barsample}) follow a similar morphology distribution, so we do not find a systematic preference of morphology for barred galaxies.  We refer to \cite{huertas2019,rodriguez2019,tacchella2019}  for preliminary analysis of the morphologies of the \TNG~ galaxies.   In their study, \cite{tacchella2019}  find  a reasonable agreement with a number of  observational relations for both disc and bulge-dominated galaxies that include PanStarrs and SDSS low-redshift galaxies (note, however, that their morphological classification gives slightly larger bulge  fractions than those found here).

\subsection{Bar sample} 
\label{subsec:barsample}
\begin{figure*}    
\includegraphics[width=2\columnwidth]{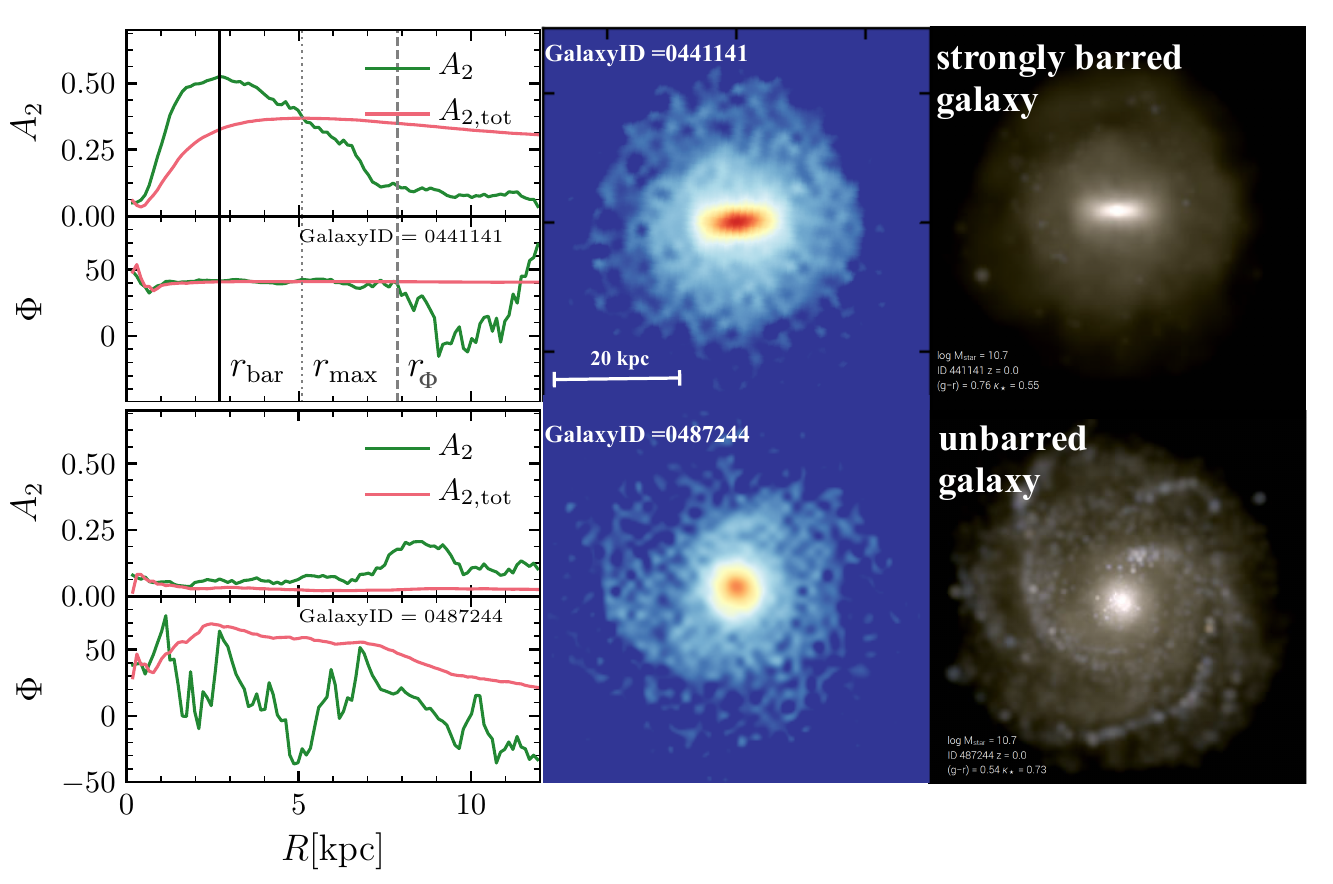}  
\caption{An example of a strongly barred and an unbarred galaxy with the same stellar mass ($\sim 10^{10.7}\Msun$) in the TNG100 simulation. \textit{Top left:} In the upper panel, the  $\Ato$ profile of the Fourier decomposition of the face-on stellar surface density (green curve) and its cumulative distribution,  $\Atot$. In the lower panel, the corresponding profiles of the phase, as defined in equations~\ref{eq:a2} and \ref{eq:phase}.  The vertical solid line indicates the location of the peak of $\Ato$ and in this work, it will be used as a proxy for the bar length ($\rbar$).  For reference,  the location of the peak of $\Atot$ ($\rmax$, the vertical dotted line) and the maximum radius where the phase of $\Ato$ remains constant  ($r_{\Phi}$, the vertical dashed line) are included.  \textit{Bottom left:} Similar to Top left panel but for an unbarred galaxy.   \textit{ Middle panels} correspond to the face-on stellar surface density map of $60\times60\times4$ kpc centred on each barred and unbarred galaxy. \textit{Top and Bottom right:} mocked images in JWST  NIRCam F200W, F115W, and F070W filters (face-on) calculated in \protect \cite{nelson2018a}. The NIRCam blue channel  highlights the young population of the galaxy and the NIRCam red channel older populations. The images illustrate the differences in the stellar population between the strongly barred galaxy (quenched with  sSFR$\sim 10^{-12}\,\rm yr^{-1}$) and the unbarred galaxy (still forming stars with  sSFR$\sim 10^{-11}\,\rm yr^{-1}$).}
\label{fig:barimage}
\end{figure*}

We identify stellar bars by Fourier decomposing the face-on stellar surface density  (e.g., \citealt{athanassoula2002}). We calculate $\Ato$ as the ratio  between the second  and zero terms of the Fourier expansion:
\begin{equation}
A_{2}(R)= \frac{|\Sigma_{j}m_je^{2i\theta_j}|}{\Sigma_{j}m_j},
\label{eq:a2}
\end{equation}
where  $m_j$ is the mass of the jth particle and $\theta_j$ is the angular coordinate on the galactic plane. The summation is performed over stellar particles within a cylindrical shell of radius $R$, coaxial to the centre of the galaxy,  and width ${\rm d}R$ that in this case is $0.12 \,\rm kpc$. We take as a proxy for the strength of the bar the maximum of $\Ato$ ($\Amax$) and the location at which the maximum is reached is assumed to be the length of the bar ($\rbar$).
We also ensure that the phase of the $m=2$ mode, $\Phi(R)$, is constant within the extent of the bar, with $\Phi(R)$ defined as:  
\begin{equation}
\Phi(R)= \frac{1}{2}{\rm arctan}\bigg[\frac{\Sigma_{j}m_j{\rm sin}(2\theta_j)}{\Sigma_{j}m_j\rm{cos}(2\theta_j) }  \bigg],
\label{eq:phase}
\end{equation}

where the summation is done in coaxial cylinders with radius $R$ and width $\rm {d} R$ from the centre of the galaxy.  
In Fig.~\ref{fig:barimage} we present an example of the radial profiles of $\Ato$ and its phase for a strongly barred and an unbarred galaxy. The cumulative profiles of $\Ato$ ($\Atot=\Ato(<R)$) and its phase, are also shown for reference. Note that $\Atot$ could be smaller than $\Ato$ because both are normalised by the zeroth term of the Fourier expansion.  We also show, for reference, the radii at which $\Ato$ peaks ($r_{\rm max}$) and within which the phase of $\Ato$ remains constant ($r_{\Phi}$). These have been used in the literature as alternative proxies for the bar length.  Note that the length of these alternative definitions could be larger than $\rbar$. Because we are interested in the evolution of the gas inside the bar, we choose $\rbar$ as our proxy for the bar extent, since it has a good constraint on the region where the bar could produce its highest effect. Nonetheless, our conclusions do not vary when choosing a different definition for the bar extent.

Clearly, the Fourier decomposition of the face-on stellar densities of the two galaxies shown in Fig.~\ref{fig:barimage} are extremely different, with the galaxy on top featuring a typical $\Ato$ profile of strongly barred galaxies, while the one shown in the bottom  does not exhibit any signature of a bar structure. The differences in the inner stellar structures of the two galaxies can be visually appreciated in the middle and right panels. Here, we show, respectively, the face-on stellar densities maps of the two galaxies and mock images from \cite*{nelson2018a} calculated for the NIRCam f200W, f115W, and F070W filters (dust is not included).    

We apply the above Fourier analysis to all the galaxies in the parent disc-galaxy sample. We select all galaxies with bar features, defining a galaxy as  barred  if $\Amax\geq0.2$, $\rbar>1\kpc$ and $\Phi (< \rbar)=\rm const$. The second condition ($\rbar>1\,\kpc$) ensures that bars are not affected by the gravitational softening length of the stellar particles  ($\sim 700 \, \rm pc$). Above this scale, we are also confident that the simulation captures the full dynamics of the interstellar medium (ISM).

We then divide the parent disc-galaxy sample into three subsamples:
\begin{itemize}
\item {\bf strong bar sample:} disc galaxies with $A_{2,\rm max}\geq0.3$
\item {\bf weak bar sample:}  disc galaxies with $0.2 \leq A_{2,\rm max}<0.3$
\item{\bf unbarred sample:} all remaining disc galaxies. 
\end{itemize}

In total, we identify 107 barred galaxies of which 59 have a strong bar and 48 have a weak bar, being  57 and 24 centrals respectively. In the remainder of the analysis, we do not distinguish between central and satellite galaxies. 
We note that almost half of the weak bars were recently formed ($\sim 0.16\, \Gyr$, i.e., one output before the end of the simulation). Thus, we can not asses  whether they are a stable or transient feature.

\begin{figure}	
\includegraphics[width=1.0\columnwidth]{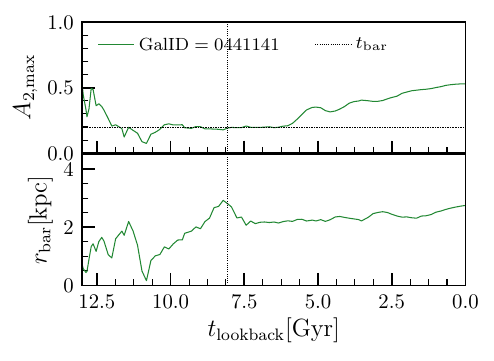}  
\caption{ The evolution of the strong bar shown in Fig.~\ref{fig:barimage} hosted in a TNG100 disc galaxy. After the formation of the bar ($t_{\rm lookback}=\tbar$, vertical dotted line), its strength (top panel) increases with decreasing $t_{\rm lookback}$  whereas its length (bottom panel) remains roughly constant as decreases $t_{\rm lookback}$.} 
\label{fig:barevol}
\end{figure}

While the spatial and temporal resolution of the \TNG~ simulation do not allow us to study in full details of the onset of the dynamical instabilities that lead to the formation of a bar, we can still trace the evolution of the bar strength and length back in time. This allows us to determine when the bar begins to be a stable dynamical feature in the host galaxy. While  some fluctuations due to disturbances and minor mergers do happen,  we find that  the strength and length of the bars increase smoothly with time.  Figure \ref{fig:barevol} shows an example  of the evolution of the bar strength and length for a typical strong bar. The vertical dotted line  indicates the lookback time chosen as a proxy for the bar age. Specifically, we define the bar age, $\tbar$, as the lookback time ($t_{\rm lookback}$) at which the two following conditions are satisfied: 
\begin{enumerate}%
\item $\Amax(t_{\rm lookback})\geq 0.2$  for $t_{\rm lookback}<\tbar$  \\ 
\item  $ | (\Amax(\tbar)-\Amax(\tbar-\Delta t))/ \Ato(\tbar)|<0.4 $,  where $\Delta t$  corresponds to the time elapsed between the output when the bar forms and two previous outputs of the simulation. $\Delta t$ takes values of $\sim 300$ Myrs.
\end{enumerate}
The first condition ensures that the bar is a stable feature, while the second one helps to set the time at which the bar is stable and not anymore a transient component subject to strong fluctuations.

\subsection{Defining $\tnorm$}
\label{subsec:tnorm}

To aid in establishing the potential effect of bars in the evolution of their host galaxies in section~\ref{sec:evolution}, we introduce $\tnorm$ defined as the normalised time since the bar formation time i.e.,
\begin{equation}
t_{\rm norm}= (\tbar -t_{\rm lookback})/\tbar,
\label{eq:tnorm}
\end{equation}
where $\tbar$ is the bar age (defined above) and $t_{\rm lookback}$ the lookback time. This time definition allows us to highlight the bar effects on the properties of their host galaxy. $\tnorm=0$  corresponds to  $t_{\rm lookback}=\tbar$, whereas  
$\tnorm=1$ corresponds to $z=0 $ ($t_{\rm lookback}=0$).  Negative values of $\tnorm$ correspond to times prior to the appearance of the bar and positive values of $\tnorm$ refer to times after the bar is already formed. 

\subsection{Constructing a control sample of unbarred galaxies}
\label{subsec:controsample}

\begin{figure}	
\includegraphics[width=\columnwidth]{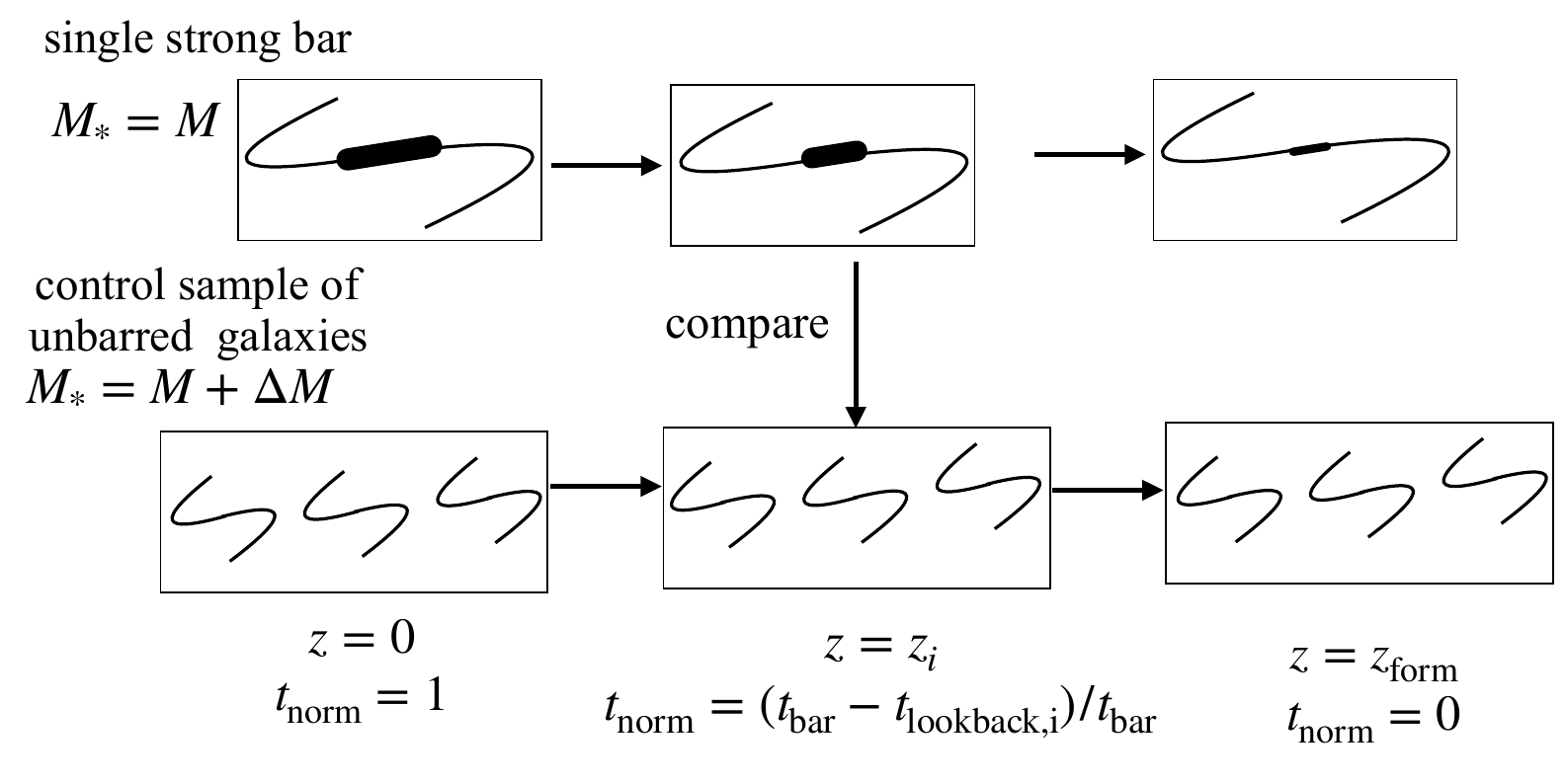}  
\caption{Representation of the construction of the control sample of unbarred galaxies at $z=0$. For each $z=0$ strongly barred galaxy with a stellar mass $M$, we select a sample of unbarred galaxies with a stellar mass between $M-\Delta M$ and $M+\Delta M$ at $z=0$. We compare the properties of the main progenitors of the control sample  to each strongly barred galaxy for the same values of $\tnorm$ where $\tnorm$ is calculated for each single strongly barred galaxy (see eq. \ref{eq:tnorm}).} 
\label{fig:controlsample}
\end{figure}

To determine the significance of bars in the evolution of disc galaxies in section~\ref{sec:evolution}, we also require a control sample of unbarred disc galaxies. The control galaxies are selected  to have similar masses to barred galaxies at $z=0$ and to be unbarred ($\Amax\leq 0.2$). To match barred to unbarred galaxies, we consider three stellar mass bins at $z=0$, from $10^{10.4-11}\Msun$ and with a $0.2$ dex width in $M_{*}$.
When we track the evolution of the barred galaxies, we also  study the history of the control sample. To do so, for any given property analysed for the progenitors of barred galaxies,  we calculate the median value of this property for the progenitors of the control sample  at all the times considered, as shown in Fig.~\ref{fig:controlsample}.  When we analyse a given property inside the bar extent, we also calculate the median value of this property for the control sample inside a central region with the same extent that all bars and at all the times examined.

\section{The population of barred galaxies at $z=0$}
\label{sec:barpropz0}
In this section, we present  the properties of the bar sample at $z=0$. We start with global properties of the bars themselves, and then move on to the properties of the host galaxies.

\subsection{Properties of the bar structures}
\label{sec:barfracstar}

\begin{figure}	
\begin{tabular}{c}
\includegraphics[width=\columnwidth]{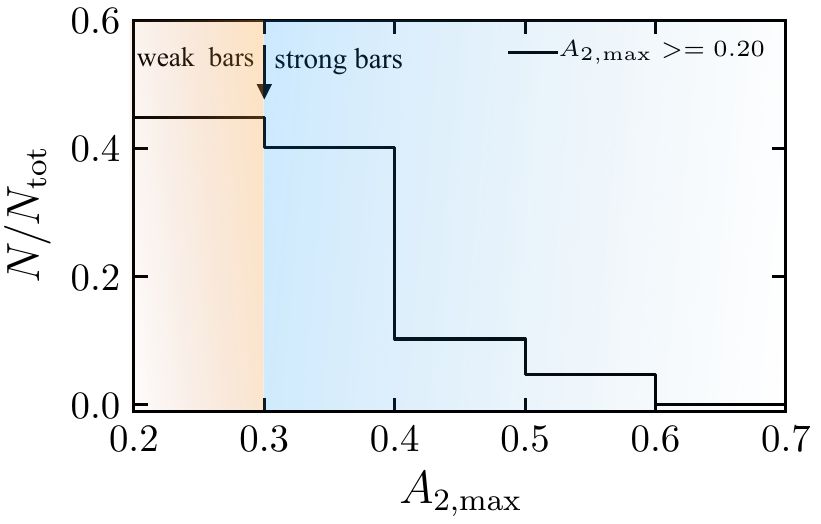}  \\
\includegraphics[width=\columnwidth]{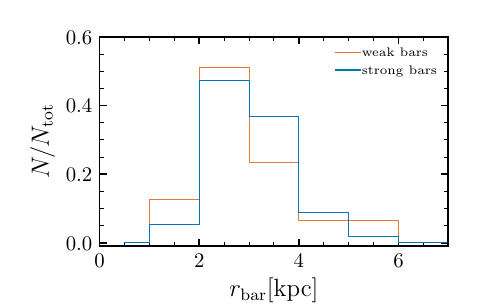}  \\
 \end{tabular} 
\caption{ \textit{Top panel}: The bar strength distribution for the parent disc-dominated galaxies in TNG100. \textit{Bottom panel:} The distribution of the bar lengths, $r_{\rm bar}$ (solid lines),  for   weak and strong bar samples.}
\label{fig:bardistribution}
\end{figure}

In Fig.~\ref{fig:bardistribution} we show the distribution of bar strengths and lengths of our  sample of barred galaxies. Most of TNG galaxies have bar strengths between 0.2 and 0.4 with a median of $\sim 0.3$. As noted above, we consider as strong bars all the structures with $\Amax\geq0.3$. The bar lengths span between $1$ and  $6$ kpc, with a median of $3$ kpc\footnote{The steep decrease at small radii, could be due to the limited resolution of the run. The simulation may not allow for the formation of sub-kpc scale bars.}. We find no significant difference between the lengths of strong bars and weak bars.
In subsection \ref{subsec:barsample}, we find that $40$ per cent of our disc galaxy sample is barred, with $22$ per cent having a strong bar and $18$ per cent exhibiting a weak bar. Such bar fractions are consistent with observational estimates obtained from the SDSS, that span between $30$ per cent and $52$ per cent \citep{barazza2008,nair2010}. These fractions at $z=0$ are also broadly consistent with the theoretical works of  \cite*{algorry2017} and  \cite*{peschken2019} who analyse the EAGLE simulation and the Illustris simulation, respectively. \cite*{peschken2019} find a lower bar fraction ( $\sim 20$ per cent) whereas \cite*{algorry2017} find a similar bar fraction ($\sim 40$ per cent) in a stellar mass range slightly higher than the one used in this work.  

\begin{figure*}	
\begin{tabular}{cc}
\includegraphics[width=\columnwidth]{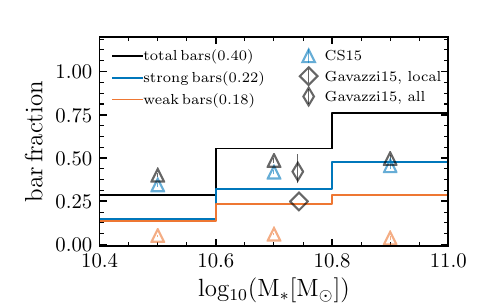} &
\includegraphics[width=\columnwidth]{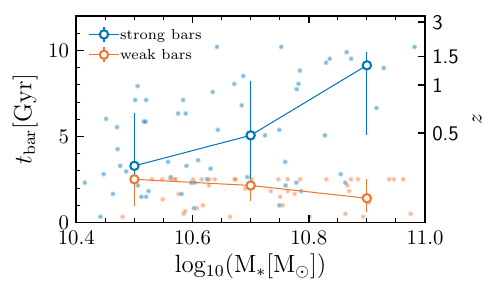} 
 \end{tabular} 
\caption{ \textit{Left  panel}: The bar fraction as a function of stellar mass for the barred galaxy sample in the TNG100 simulation, and separated into  weak and strong bars. The markers show observational estimates from \protect \cite*{cervantes2015} (CS15, triangles), who use late-type galaxies in the SDSS-DR7 and \protect \cite*{gavazzi2015} (diamonds) who used a sample of star-forming galaxies from ALFALFA in the local and Coma superclusters. The numbers in brackets correspond to the total bar fraction in each sample. The bar fraction increases with stellar mass. \textit{Right panel:} The median bar age for the strong bar and weak bar samples. Error bars represent the 20$^{\rm th}$ and 80$^{\rm th}$ percentiles of the distribution and the dots represent the individual galaxies. Strong bar ages  increase with increasing stellar mass, although with a large scatter. Weak bars have formed recently.} 
\label{fig:galfracstarmass}
\end{figure*}

Observationally, there seem to be indications of a relation between the bar fraction and stellar mass. \cite{cervantes2015}, for example, present a sample of late-type galaxies from the SDSS-DR7 where bars are detected by visual inspection. They find an increasing trend in the bar fraction with stellar mass. Also \cite{gavazzi2015}, using a sample of star-forming galaxies from ALFALFA  in the regions of the Local and Coma superclusters, find an abrupt increase in the strong bars with mass for visually identified strong bars. The top left panel of Fig.~\ref{fig:galfracstarmass} shows the bar fraction as a function of stellar mass. We find an increase in the bar fraction with increasing stellar mass. This relation is more clear for strongly barred galaxies and is in rough agreement with the observational data. However, a more detailed comparison is needed that accounts for potential effects of galaxy selection, galaxy morphology criteria, and methods of bar identification for each observational dataset and the simulation.


The increasing trend in the bar fraction with  stellar mass (at least for the strong bars)  is expected if the discs in massive galaxies become dynamically cold  earlier than those in less massive galaxies, as discussed by \cite{sheth2012}. 
There, the authors analyse a sample of disc galaxies in the COSMOS survey and find that the bar fraction declines with increasing redshift from $z=0$  to $z=0.84$. This change is larger in their lowest mass galaxies ($M_{*}\sim10^{10}\Msun$). They  suggest that  lower mass galaxies may not form bars because they could be dynamically hotter than more massive systems, due to  differences in the assembly histories.  In line with this interpretation, we find a positive correlation between bar age and stellar mass for the strong bar sample, with bars in more massive galaxies being older (see right panel of Fig.~\ref{fig:galfracstarmass}). To make sense of the bar ages, more than $10$ per cent of the strong bars are already in place by $z\approx1.5$ ($10$ Gyr ago) and more than  $\approx 50$  per cent of them already formed by $z\sim0.5$ ($3.75$ Gyrs ago). Instead, bars in smaller galaxies have all formed after $z=1$. In the case of weak bars,  we find that they formed later with a median age of $\sim 2.3$ Gyr ($z=0.2$).  

 
\begin{figure}    
\includegraphics[width=1.0\columnwidth]{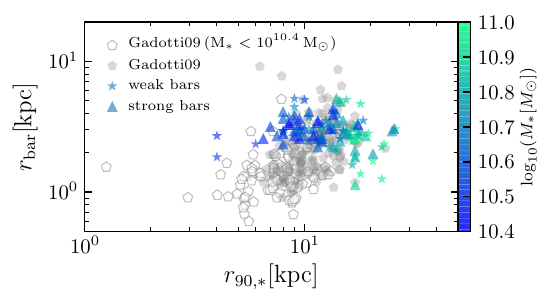} 
\caption{The bar lengths in TNG100 barred galaxies as a function of $r_{90,*}$ where $r_{90,*}$ is the radius containing $90$ per cent of the total light. Symbols are colour-coded according to the stellar mass of the host galaxy. Stars represent weak bars and triangles strong bars. Observational estimates from \protect\cite*{gadotti2009} for disc galaxies with a stellar mass larger than $10^{10.4}\Msun$ are included as grey filled hexagons. Observed disc galaxies below this threshold correspond to grey empty hexagons.}
\label{fig:gassize}
\end{figure}

The extent of the bar also seems to be connected to the host galaxy, in that there is a relation between the bar length and the size of the galaxy.  \cite{gadotti2011}, for example, analyses a sample of 300 barred galaxies in the local universe and studies the relation between bar length  (defined as either the radius that contains half of the light coming from the bar, or as the bar semi-major axis)  and galaxy size (defined as either the disc scale length $h$ or $r_{90,*}$, which is the radius containing 90 per cent of the total light). 
The author finds that the bar length correlates with the galaxy size independently of the definitions used. In Fig.~\ref{fig:gassize} we show the predictions of \TNG~ for this relation using $r_{90,*}$ and $\rbar$ and how our results compare with the observations. Overall, we see that the sizes of our bars are consistent with the ones from \cite{gadotti2011}, although we do not  find a significant correlation with galaxy size (regardless of the definition of $\rbar$ that we use). We find that the range of galaxy masses, sizes and bar extents is wider  in the observed barred galaxy sample ($M_{*}\geq10^{10}\Msun$, $r_{90,*}=1-23$ kpc and $\rbar=0.6-9.17$ kpc) than used in this work ($M_{*}\geq10^{10.4}\Msun$, $r_{90,*}=4-26$ kpc and $\rbar=1-6$ kpc).
Interestingly, \cite*{algorry2017}, using a barred sample from the EAGLE simulation,
also did not find a correlation between the bar length with galaxy size. Such a lack of significant correlation in both simulations could be a consequence of numerical resolution. In future work, we will explore this issue in more detail using the higher resolution Illustris simulation \TNGF.

\subsection{ Properties of barred galaxies}
\label{sec:bar_galaxies}

After their buildup, bars are expected to influence the evolution of their host galaxies. It is believed that  bars play an important role  in the redistribution of stars and gas (e.g., \citealt{athanassoula1992}) and  in  the star formation of galaxies (e.g., \citealt{gavazzi2015}). Moreover, bars could be responsible for AGN triggering, although it has been suggested that most of the AGN activity could take place during the bar formation when a clear strong bar is hardly observable (see the discussion in \citealt{fanali2015} and references therein).  
 
Upon formation, it is expected that bars interact with the gas disc producing a net torque between the bar and the gas. As a consequence,  an exchange of angular momentum and energy takes place between them.  It is expected that cold gas is more prone to be affected by the bar. If so,  the net torque could funnel gas into the central part of the galaxy, producing a rapid star formation burst, and possibly also feeding the central supermassive black hole. As a consequence, disc galaxies with strong bars will burn their gas reservoirs more rapidly than their unbarred counterparts.  This should be reflected in the properties of the gas, especially in the specific star formation rate-stellar mass diagram.

In this section, we examine the properties of the galaxies hosting bars, comparing them with the ones of unbarred systems, to see if we can get some hints on the role of bars in driving galaxy evolution. 

In addition to stellar mass, one of the most fundamental properties of a galaxy is its level of star formation.
Fig.~\ref{fig:barimage} shows the stellar light mock images of a barred and an unbarred galaxy with similar stellar ($\sim 10^{10.7}\Msun$) and black hole masses ($\sim 10^{8.2}\Msun$).
The blue channel from NIRCam (F070W) in the mock images emphasises the young population whereas the red channel (F200W and F115W) the older stellar population. 
 Just by visually inspecting the mock images, it is clear that the two galaxies have different stellar populations.   The unbarred galaxy presents a high level of star formation, with clumpy star-forming regions 
distributed all across the galaxy and spiral arms whereas  the barred galaxy is closer to quiescence.

\begin{figure*}
\begin{tabular}{cc}

\includegraphics[width=1.\columnwidth]{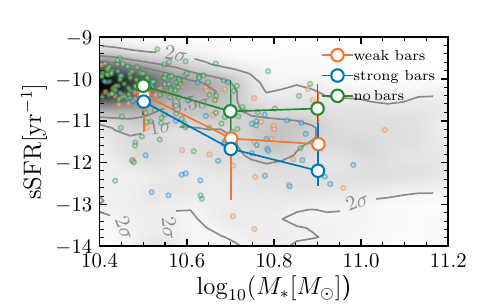} &
\includegraphics[width=1.0\columnwidth]{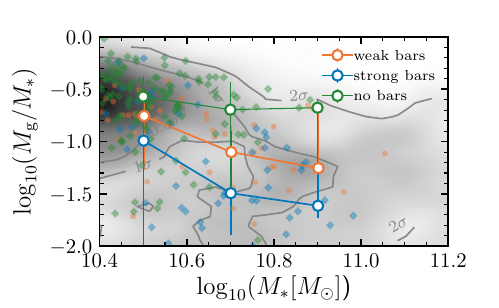}\\

\end{tabular}
 \caption{\textit{Left panel:} The sSFR as a function of stellar mass for TNG100 galaxies. The solid lines and markers represent the median relation for weak, strong and no bars, as specified in the legend. Scattered symbols correspond  to each galaxy. Grey contours and the diffuse density map represent all the galaxies in the TNG100 simulation. The figure highlights the differences  in SF activity between barred and unbarred galaxies. \textit{Right panel:}  The gas-to-stellar mass ratio  as a function of stellar mass for our three different samples. For a given stellar mass bin, galaxies with strong and weak bars present less gas than  unbarred  galaxies  overall.} 
\label{fig:ssfrdiagram}
\end{figure*}

In the left panel of Fig.~\ref{fig:ssfrdiagram} we show the specific star formation rate (sSFR)-stellar mass plane for weakly, strongly, and unbarred disc galaxies  with the sSFR defined as the instantaneous star formation rate over stellar mass inside $2\rhalf$. The solid lines with thick markers and error bars represent the median of each sample and the 20$^{\rm th}$ and 80$^{\rm th}$ percentiles respectively. The scattered symbols represent each galaxy in its sample. We include as contours star-forming galaxies in the simulation with a stellar mass larger than $10^{10.4}\Msun$ and sSFR$\geq 10^{-14}\rm yr^{-1}$. The figure shows that  at lower stellar masses ($\lsim10^{10.6}\Msun$) most of the galaxies are  star-forming. However, for larger stellar masses, the barred disc galaxies (blue and orange lines) present sSFRs that approach the typical values of quenched systems, also in comparison with the unbarred disc sample. Overall,  unbarred  galaxies present sSFR$\sim 10^{-10.3}\rm \, yr^{-1}$ and strongly barred galaxies sSFR$\sim 10^{-11.7}\rm \, yr^{-1}$. 

 It is worth mentioning that the IllustrisTNG simulation roughly reproduces many important observables for the current investigation in the local Universe such as galaxy sizes \citep{genel2018}, the star formation rate density across time and the galaxy stellar mass function at $z=0$ \citep{pillepich2018a}. In particular, \cite{nelson2018a} show that the simulation reproduces the galaxy colour bi-modality of blue galaxies and red galaxies observed  at low redshift and compare it to the observed galaxy distribution from the SDSS.

The drop in star formation in massive disc galaxies with a bar is linked to a drop in the gas content of the barred galaxies. The right panel of Fig.~ \ref{fig:ssfrdiagram} shows the median gas fraction ($M_{g}/M_{*}$) as a function of stellar mass for disc galaxies with no, strong, and weak bars.   The gas-to-stellar mass ratio is about $1/3$ for the unbarred disc galaxies  (green line), and it is fairly constant with stellar mass. On the other side,   the strongly- and weak-barred galaxies present much lower gas-to-stellar mass ratios, and the fractions decrease with increasing stellar mass, becoming up to more than 1 dex lower than those presented by their unbarred counterparts in the most massive galaxies.

The lower gas-to-stellar mass ratio in strongly barred galaxies is in rough agreement with observational estimates. For instance,  \cite{cervantes2017}, using a barred galaxy sample selected from SDSS and studied with the ALFALFA survey, find that the fraction of strong bars increases with decreasing HI-to-stellar mass ratio for a given stellar mass.  Similarly, \cite{masters2012} point out a residual anti-correlation between the bar fraction and the HI-to-stellar mass ratio using Galaxy Zoo combined with the ALFALFA survey, after accounting for dependencies with galaxy stellar mass and colour.

The fact that bars in the local Universe are more likely to be hosted by galaxies with low gas-to-stellar ratios and quenched star formation  has been interpreted in the literature under two different scenarios: (1) the bar is most likely to form and to rapidly grow  in disc galaxies that are gas-poor (e.g., \citealt{athanassoula2013}) or (2) the bar  has a strong role in redistributing gas within the galaxy, producing  bursts of star formation and finally leaving the host galaxy quenched and gas-poor (e.g., \citealt{cheung2013,gavazzi2015}). To explore these two scenarios, in the following section we study the evolution of galaxies hosting strong bars across cosmic time and compare them to the evolution of a control sample of unbarred galaxies.

\section{ The evolution of the strongly barred galaxies}
\label{sec:evolution}

In this section, we explore the cosmological evolution of strongly barred and unbarred disc galaxies at $z=0$, focusing on both the bar properties and the galaxy properties. We exclude weak bars since, as we have seen earlier, a large fraction of such structures are not fully assembled and their properties might be affected by the limited resolution of the simulation. We start with a qualitative exploration, by inspecting visually the evolution of a strongly barred and an unbarred galaxy. We then look into the evolution of the strength and length of the strong bars and the connection with the star formation rate of the host galaxies.  We move to a more detailed analysis of the evolution of the galaxy properties, finishing with a study of the merger histories of the two classes of galaxies.

\begin{figure*}  
\centering
\includegraphics[width=2\columnwidth]{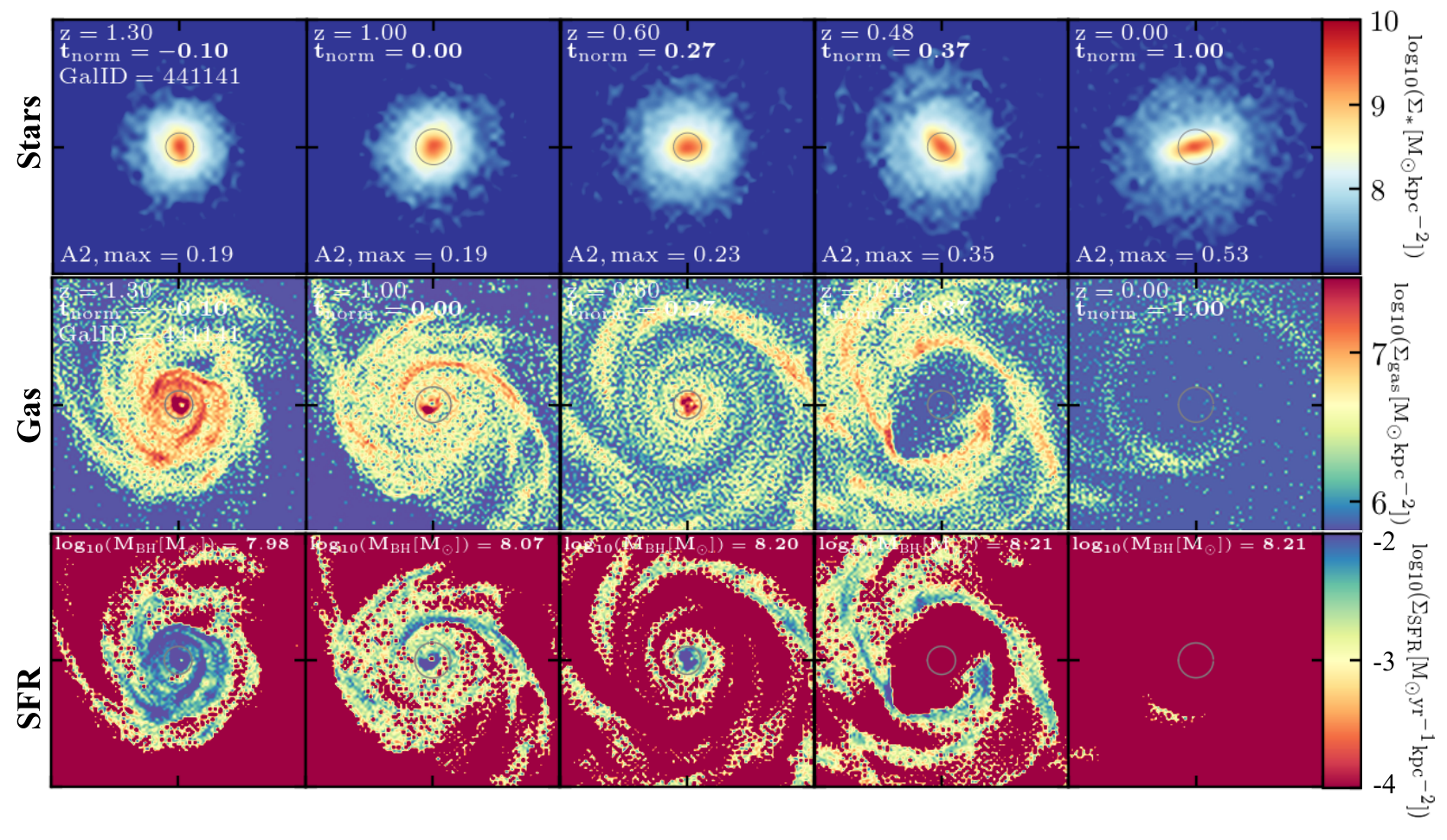}  
\caption{The evolution of a strongly barred galaxy with a final stellar mass $\sim 10^{10.7}\Msun$ at $z=0$ in the TNG100 simulation. Each row shows the time evolution of different galaxy components via face-on surface density maps of stars (top panels),  gas  (middle panels) and  instantaneous star formation rate, SFR (lower panels). Maps are obtained from slices of $40 \times 40 \times 4 ~\kpc$. The bar is in place by $z\sim 1$ ($\tnorm =0$, second column), and keeps growing in length and strength. The central circle indicates the bar length, defined to be the maximum of $\Ato$ at each time. At times before the epoch of bar formation, the central circle has a radius of $2\kpc$.   As the bar strength increases,  the gas content in the central part decreases and so does the SFR. By $z=0$, the complete galaxy becomes passive (sSFR$\sim 10^{12} \, \rm yr^{-1}$).}
\label{fig:strongbarevol}
\vspace*{\floatsep}
\includegraphics[width=2\columnwidth]{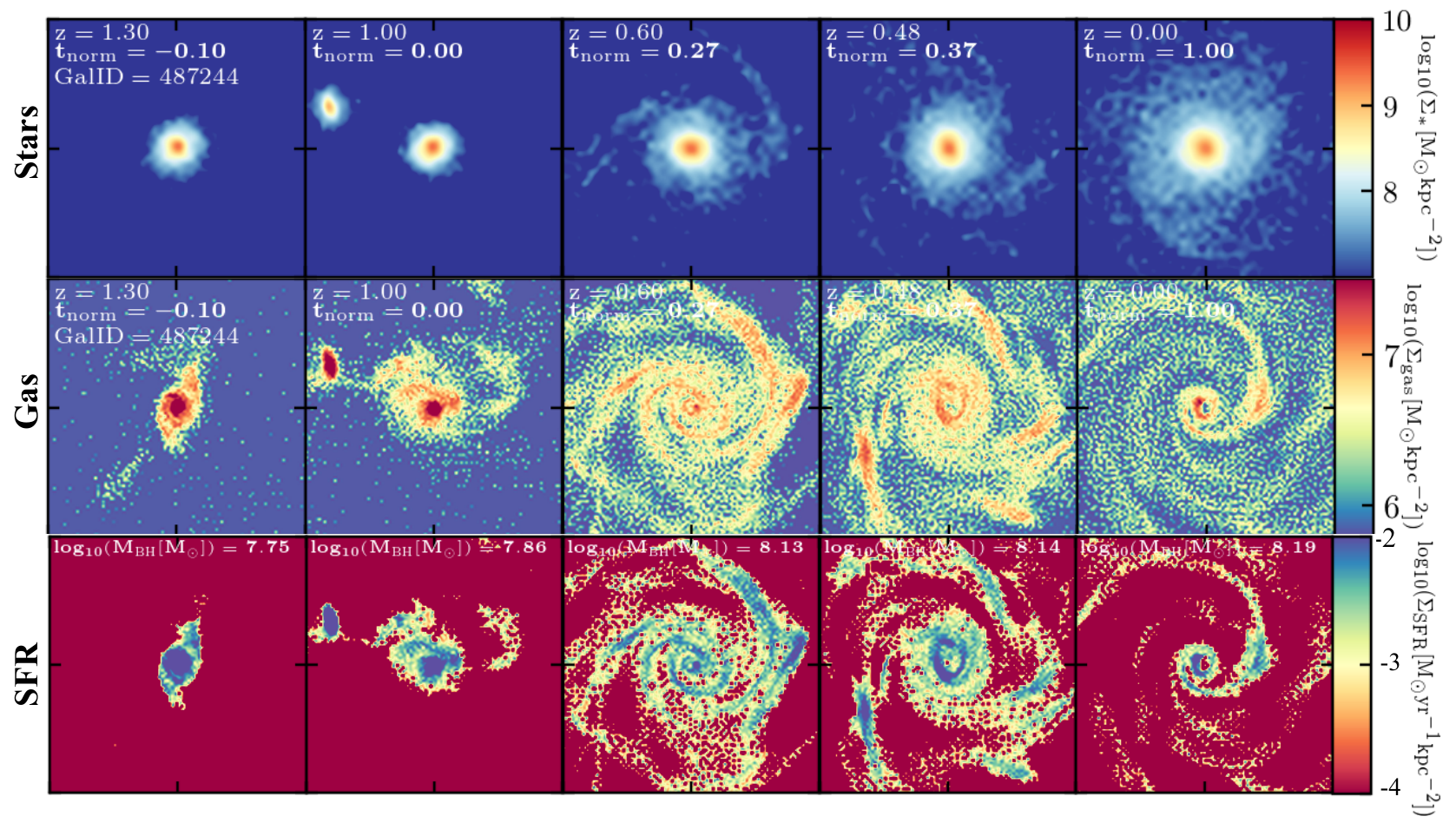}  
\caption{The evolution of an unbarred galaxy at $z=0$ in the TNG100 simulation. The stellar mass at $z=0$  is similar to the one from strongly barred galaxy shown in Fig.~\ref{fig:strongbarevol}. Maps show the stellar component, the gas content and the SFR of the galaxy at the same snapshots as in Fig.~\ref{fig:strongbarevol}.  Contrary to the strongly barred galaxy, this galaxy features a very small disc at high redshift and experiences a minor merger at $z=1$.  The stellar mass increases significantly with time and the gas content decreases  at lower redshifts in comparison with the strongly barred galaxy. By $z=0$, the galaxy  continues forming stars. } 
\label{fig:nonbarevol}
\end{figure*}
\subsection{Case study: The evolution of a strongly barred and an unbarred galaxy}
\label{subsec:casestudy}

To have a sense of the differences in the evolution between barred and unbarred galaxies, we start with a qualitative exploration of the properties of two representative disc galaxies with a similar stellar mass and a black hole mass at $z=0$.  Fig.~\ref{fig:strongbarevol}  shows the face-on surface density of the stellar, the gas and the instantaneous star formation rate components at different redshifts (from $z=1.35$ to $z=0$) for the barred galaxy already shown in Fig.~\ref{fig:barimage}. 
From the $\Ato$ analysis at different snapshots (see Fig.~\ref{fig:barevol}), we determine that the bar of this galaxy is already in place at  $z\sim 1$ (labelled as $\tnorm=0$ in the maps). With time, the bar increases in strength, as visible in the stellar density maps (top panels), and, by $z\sim 0.5$, the bar has a $\Amax>0.3$, becoming one of the strongest bars in the simulation.    The central circles in the maps correspond to $2\,\kpc$ at times prior to the bar formation ($\tnorm<0$) whereas at subsequent times, circles correspond to the bar extent. Regarding the gas component (middle panels),  the gaseous disc seems to be well defined at early times, with a high concentration of gas in the central part. Correspondingly, the galaxy displays high levels of star formation (bottom panels), mainly concentrated in the central region and in the spiral arms. Once the bar is fully settled, the gas content decreases, especially in the central part and, by $z=0.5$, the galaxy nucleus is depleted of gas. Star formation continues at larger radii without any abrupt change, slowly decreasing with time. By $z=0$  the entire galaxy is quenched as a consequence of the decreased cosmological inflow of pristine gas onto the already formed galaxy \citep[see also the discussion in][]{gavazzi2015}.

In contrast,  Fig.~\ref{fig:nonbarevol} shows the face-on surface densities of the stellar, the gas and the instantaneous star formation rate (SFR) components of an unbarred galaxy at the same redshifts as in Fig.~\ref{fig:strongbarevol}. The extended stellar disc of this galaxy forms at later times than the disc of the barred galaxy. The growth is fostered by a minor merger happening shortly after $z\sim 1$, as can be seen also in the gas maps (middle row). Star formation (bottom panels) becomes more significant through the disc and along the spiral arms. Eventually, star formation starts decreasing at very late times,  at a much slower rate in comparison to the strongly barred galaxy. In particular, the central region of the barred galaxy is already quenched and depleted of gas by $z\sim 0.5$ while the unbarred galaxy still contains star-forming gas at $z=0$. This qualitative analysis is already showing important differences in the history of barred and unbarred galaxies at $z=0$. \\

In the next subsections, we quantify these differences using our complete disc samples of strongly barred and unbarred galaxies.

\subsection{Evolution of bar structures}
\label{sec:evolsb}
\begin{figure*}
\includegraphics[width=2.2\columnwidth]{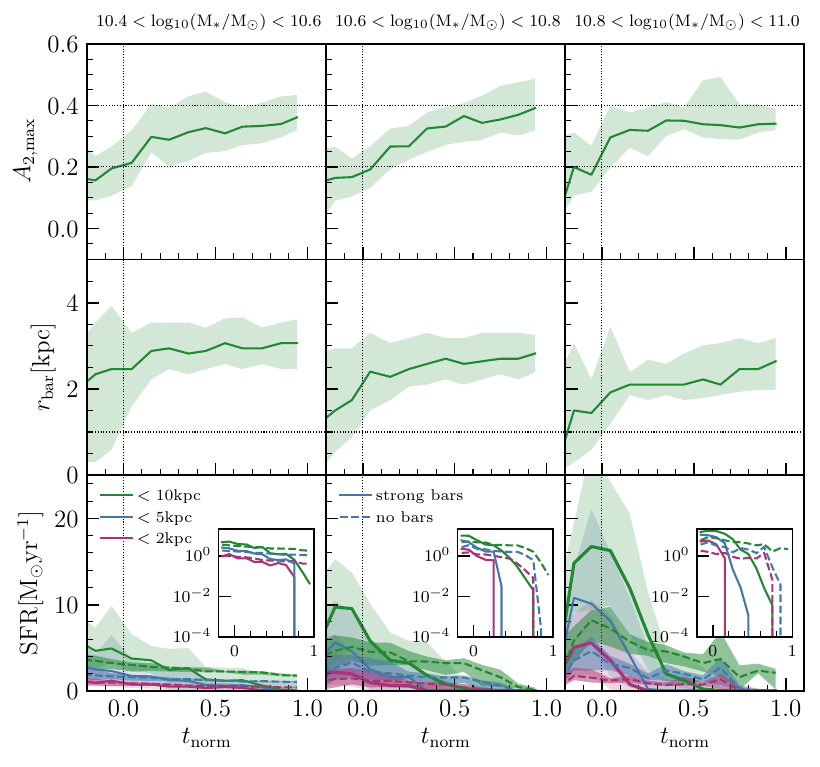}  
\caption{ \textit{Top and Middle panels:} The evolution of the bar properties in strongly barred and unbarred galaxies at $z=0$ in the TNG100 simulation, in terms of  $\tnorm$, as  defined in eq.~\ref{eq:tnorm}, in three bins of stellar mass as labelled in each column. On average,  the bar strength and  length increase with time.  \textit{Bottom panel}: The evolution of the star formation rate (SFR) at different apertures as indicated in the legend.  The solids lines indicate the median SFR for strongly barred galaxies, while the dashed lines  the one for  unbarred galaxies.  The coloured regions represent the 20$^{\rm th}$ and 80$^{\rm th}$ percentiles of each distribution. The inset plot shows the logarithmic SFR as a function of $\tnorm$.  The SFR in strongly barred galaxies, specially in their nuclear part, decreases steeply  over time once the bar is settled. This drop in SFR is more pronounced in more massive galaxies.} 
\label{fig:sb_evolution}
\end{figure*}
 \begin{figure*}
\includegraphics[width=2\columnwidth]{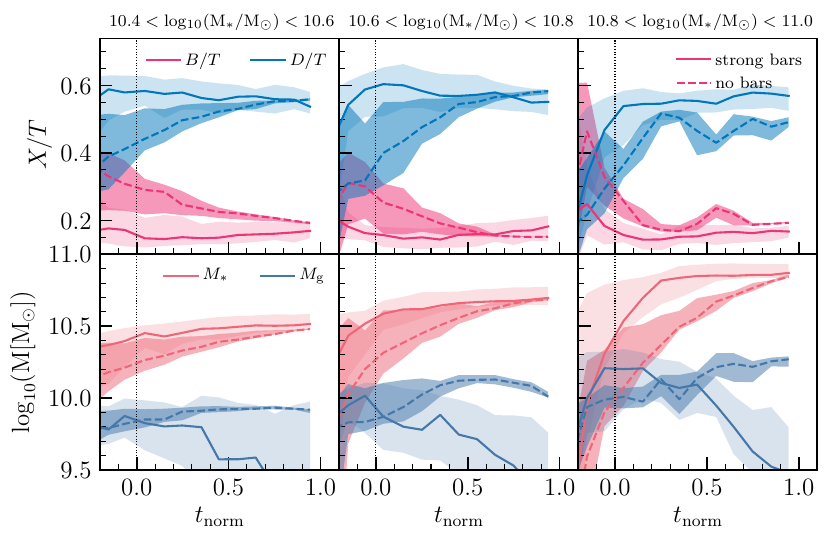} 
\caption{The evolution of properties of unbarred  and strongly barred galaxies at $z=0$ in the TNG100 simulations, in three bins of stellar mass as indicated in each column. \textit{Top row}: The evolution of  $B/T$ and $D/T$ of strongly barred (solid lines) and unbarred galaxies (dashed lines) in terms of  $\tnorm$ (eq.~\ref{eq:tnorm}).  At the epoch of bar formation  ($\tnorm =0$),  strongly barred galaxies exhibit a dominant disc component compared to the one of unbarred galaxies.  \textit{Bottom row:} The median growth of stellar and gas components within $2\rhalf$ as a function of $\tnorm$. The coloured regions represent the 20$^{\rm th}$ and 80$^{\rm th}$ percentiles of each distribution. The gas  component  decreases more steeply  in strongly barred galaxies  than the one in unbarred galaxies. Strongly barred galaxies acquired most of their stellar mass earlier than unbarred galaxies.} 
\label{fig:gal_evolution}
\end{figure*}
To determine how our results depend on galaxy mass, from now on we will analyse both strongly barred and unbarred galaxies in three bins of  stellar mass ($10^{10.4-10.6}\Msun$,  $10^{10.6-10.8}\Msun$ and $10^{10.8-11}\Msun$) that are based on the stellar mass content (within $2\rhalf$) at $z=0$.  
Given that we analyse the properties of strongly barred galaxies as a function of $\tnorm$ (the normalised time since the bar formation; eq.~\ref{eq:tnorm}), for the unbarred galaxies we will take the median properties of the galaxies at all the different times considered for the strongly barred sample (see subsection~\ref{subsec:controsample}).  
Fig.~\ref{fig:sb_evolution} shows the median evolution of the bar strength (first row) and the bar length (second row) in terms of $\tnorm$, with each column corresponding to a stellar-mass bin.  The figure shows that, overall, the bars grow in strength and length with time.  However, there is no clear relation between the bar strength and stellar mass, nor between the bar length and galaxy mass. Only the bar age seems to be mildly dependent on stellar mass, as we have shown in Fig.~\ref{fig:galfracstarmass}. Thus, the final size of the bars seems not to be correlated with the time of formation. 

The last row of Fig.~\ref{fig:sb_evolution} shows, instead, the evolution of the median star formation rate (SFR) at three different galaxy radii: 2, 5 and 10 $\rm kpcs$.   The strongly barred galaxies present high SFRs during the phases of bar formation. Once the bar is fully established, the SFR starts to decrease rapidly in time, especially for the central part of the strongly barred galaxies (purple solid lines).  This occurs in all the stellar mass bins, but the drop is more pronounced in the most massive galaxies (right column of Fig.~\ref{fig:sb_evolution}).  
We also include the SFR evolution within the same apertures for the unbarred galaxies (dashed lines in Fig.~\ref{fig:sb_evolution}). The typical SFR of unbarred galaxies shows a much different time evolution: it is lower than the one of barred galaxies at early epochs ($\tnorm<0$), but, unlike the barred galaxies,  the decrease in their star formation history is more gradual at all apertures. By $z=0$ ($\tnorm=1$), unbarred galaxies are still star-forming, while barred galaxies are quenched, as we have already shown in Fig.~\ref{fig:ssfrdiagram}. The striking differences in the evolution of the SFR for barred and unbarred galaxies seem to be consistent with a scenario of bar formation via instabilities in periods of high star formation rates.  Once the bar is settled, bar-driven quenching can operate, which we will further discuss in section~\ref{subsec:quenching}.

\subsection{Evolution of galaxy properties}
\label{sec:evolgal}

We now move to the redshift evolution of the properties of barred and unbarred galaxies, with the goals of exploring the physical conditions in which the bar develops and understanding the processes that lead to the quenching of barred galaxies. We start by analysing the morphology evolution of strongly barred and unbarred galaxies. 
Both samples by construction, have similar median disc-to-total mass $D/T$ and bulge-to-total mass $B/T$ ratios at $z=0$ (Fig.~\ref{fig:gal_evolution}, top row). The time evolution of the galaxy morphology is, however, very different in the two samples. Barred galaxies have already a well-established disc-dominated morphology by the epoch of bar formation ($\tnorm=0$). In the most massive bin of stellar mass, we see clearly how the disc is rapidly building up while the bar is settling in. Barred galaxies retain their morphology as time evolves, remaining  disc galaxies.   Unbarred galaxies, instead, present much higher  $B/T$ at early times, with much less defined disc components.   With time, unbarred galaxies undergo a  significant morphology evolution, with the disc component continuing to grow until recent times, reaching median  $D/T$ values similar to the ones of strongly barred galaxies. 

\begin{figure*}    
\includegraphics[width=2\columnwidth]{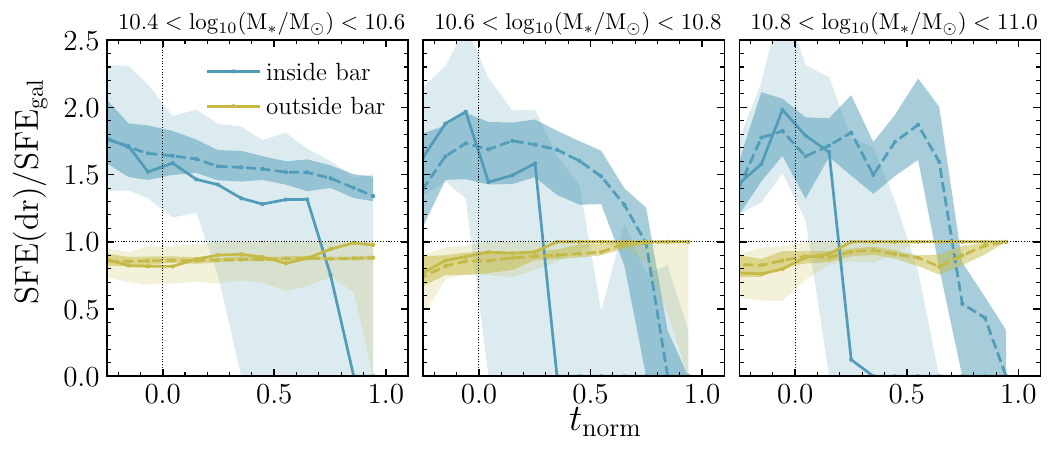}  
\caption{The median evolution of the \textit{normalised} star formation efficiency, $\SFEdr$, inside (blue lines) and outside the bar (yellow lines),  in strongly barred  (solid lines) and  unbarred galaxies (dashed lines) at $z=0$ , in three bins of stellar mass as indicated in each panel. The coloured regions represent the 20$^{\rm th}$ and 80$^{\rm th}$ percentiles of each distribution.  Strongly barred galaxies present a rapid SF quenching  as the bar evolves compared to the external parts of the galaxies and to the nuclear region of unbarred galaxies.} 
\label{fig:sfe}
\end{figure*}

The bottom row of Fig.~\ref{fig:gal_evolution}  shows the median growth of stellar (red)  and gas (dark blue) masses within $2\rhalf$ for the two galaxy samples, again for  the three  bins of stellar mass. 
We find differences in evolution between strongly barred galaxies and  unbarred galaxies. While the two samples have similar stellar masses at $z=0$ ($\tnorm=1$), by construction, the stellar content of strongly barred galaxies is significantly larger than the one of unbarred galaxies at the epoch of bar formation.  
At $\tnorm=0$,  the stellar mass in strongly barred galaxies is higher than the one in unbarred galaxies by up to $0.4$ dex. The stellar mass assembly is much faster for barred galaxies, while unbarred galaxies grow more gradually with time. 

This is consistent with the evolution of the SFR previously (see Fig.~\ref{fig:sb_evolution}), where we have shown that the typical SFR of barred galaxies is larger at early times, during bar formation, and decreases rapidly later on. Unbarred galaxies, instead, show a SFR that drops weakly with time. 
The evolution of the average gas content in the two samples of galaxies also reflects this trend.   At early times, when bars are forming, barred galaxies display a larger gas content than the unbarred control sample. Later on, at $\tnorm>0$, the mass in gas for the strongly barred galaxies significantly decreases whereas for unbarred galaxies it steadily increases with time.

\subsection{Star formation \textit{quenching} within the central disc}
\label{subsec:sfe}

In the previous sections, we have shown how the star formation rate (SFR) of barred galaxies strongly declines after bar formation, especially at small apertures, and that this trend is more striking in more massive galaxies.  We further explore the role of bars in  influencing the star formation of their host galaxies by examining the efficiency of star formation, $\SFE$. In particular, we look at the SFE inside and outside the bar extent, relative to the averaged star formation efficiency in the whole galaxy.
 The star formation efficiency  ($\SFE$) within a given $\rm dr$ and at a given $\tnorm$ is defined as: 
\begin{equation}
 \rm SFE(dr) = \rm  SFR(dr)/M_{\rm cold}(dr),
\end{equation}
where  $\rm SFR(dr)$ is the star formation rate and $M_{\rm cold}(\rm dr)$ is the total mass of the star-forming gas cells \footnote{ We consider gas cells as star-forming  when its density exceeds $n_{\rm H}=0.1 \,\rm cm^{-3}$  and the instantaneous $\rm SFR>0$; see \citealt{pillepich2018a} for details.}  within $\rm dr$. We normalise the SFE(dr) by using  $\SFE_{\rm gal}$ as a factor of normalisation, where $\SFE_{\rm gal}$ is the star formation efficiency of the galaxy using $15\,  \kpc$ outer boundary as a definition. If  $\SFEdr$  is close to 1, star formation in this given region is equally efficient as the total galaxy. If $\SFEdr$ approaches 0, instead, the given region is not forming stars while if it is above 1, star formation  is higher in this given region than the average in the galaxy.   In Fig.~\ref{fig:sfe} we show  the evolution of $\SFEdr$  in two regions: inside the bar extent  ($0 <r< \rbar(t)$)  and outside the bar extent  ($\rbar(t)<r< 15\rm kpc$ ).  For times prior to $\tbar$ ($\tnorm<0$),     
we set a fixed value of $\rbar=2 \,\rm kpc$.  To explore the differences with the unbarred sample, we also take the median $\rm SFE$ within the same bar lengths (as defined by the strongly barred sample) and at the corresponding times as described in subsection \ref{subsec:controsample}.   


From Fig.~\ref{fig:sfe} we, first of all, see that outside the bar (yellow lines), the $\SFEdr$ is close to 1 for both barred and unbarred galaxies, indicating that outside the bar the efficiency of star formation is similar to the average one within the galaxy. This seems also to be independent of time, with only some mild increase after the bar gets settled.  Inside the bar region, instead, the $\SFEdr$ decreases with time, and this is independent of the galaxy presents a bar or not. In the strong bar sample, however, the drop of $\SFEdr$ happens earlier than for the unbarred sample. In particular, the drop happens shortly after the formation of the bar in the case of the most massive galaxies\footnote{Note, however, that because the bar age is, on average, larger in more massive galaxies, the interval in physical time ($\Delta \tnorm=1$) is considerably shorter in less massive galaxies. So that, this is consistent with a fast quenching in such systems as well.}. \\
This shows a possible scenario of  \textit{bar quenching} of the galaxy central regions, as the presence of a bar seems to promote a rapid consumption of the central cold gas. We emphasise, however, that more detailed conclusions could be drawn when analysing the $\SFEdr$ at different apertures within the bar. As we will further discuss later, we plan to do so in future work, using the higher resolution of  IllustrisTNG50. In section~\ref{subsec:quenching} we will also discuss the possibility that gas could be heated or expelled by other physical processes, such as  AGN feedback. 

\subsection{Merger histories}
\label{subsec:mergerhistories}
\begin{figure}    
\includegraphics[width=1\columnwidth]{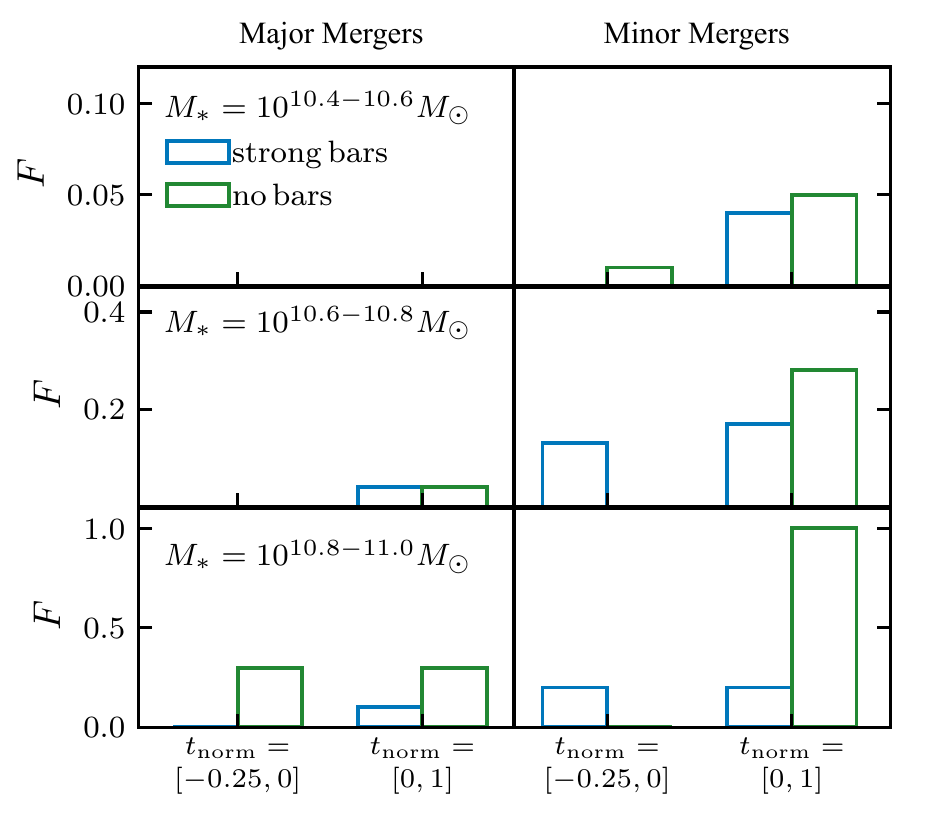} 
\caption{Mergers histories of $z=0$ strongly barred (blue lines) and unbarred galaxies (green lines). The left part of the plot shows the fraction of galaxies that experience at least one major merger at times prior to bar formation ($-0.25<\tnorm<0$, left histograms) and at times after  bar formation  ($\tnorm\geq0$, right histograms). The right part of the plot, instead, shows the  fraction of galaxies that experiences one or more minor mergers. Strongly barred galaxies present similar merger histories than their unbarred counterparts dominated by minor mergers. }
\label{fig:mergers}
\end{figure}

To further explore differences in the evolution of strongly barred and unbarred galaxies, we trace the history of each galaxy and count the number of major and minor mergers experienced both before ($\tnorm<0$) and after ($\tnorm>0$) bar formation.  We classify  major mergers when  the stellar mass ratio between the secondary and the primary galaxy, $f_{*}$, is $\geq 0.1$, while minor mergers are defined for mass ratios in the range $0.01\leq f_{*}<0.1$.
Masses of the merging galaxies are estimated at the snapshot prior to the one the SubLink algorithm considers the galaxies already merged\footnote{We also consider the stellar mass ratios using the masses at the time when the secondary galaxy reaches its maximum stellar mass in the past.  Overall, we find a small increase in the fraction of galaxies that experience one or more merger, but the main differences between barred and unbarred galaxies are preserved.}.  To ensure that structures have been properly followed, we only consider mergers where both galaxies have a  stellar mass larger than $10^{9}\Msun$ ($\sim1000$ times the initial mass of gas cells).

Fig.~\ref{fig:mergers} summarises these results: both strongly barred and unbarred galaxies show similar quiet
merger histories with a null or a small fraction of galaxies that experience at least a major merger before and after the bar appears (left panels).  We note that the most massive sample ($M_{*} =10^{10.8-11}\Msun$, bottom left panel)  exhibits the highest fraction of unbarred galaxies that experience at least one major merger  (30  per cent before and after the bar forms) whereas strongly barred galaxies have a lower fraction (0, 10 per cent before and after the bar forms respectively).

 The right panels of Fig.~\ref{fig:mergers} also show the fraction of galaxies 
that experience at least one minor merger at times prior to the epoch of bar formation.  There is no clear difference between strongly barred and unbarred galaxies.  For instance, 1 per cent of unbarred galaxies with $M_{*} =10^{10.4-10.6}\Msun$, experience at least one minor merger against a null fraction of strongly barred galaxies. Oppositely,   in the stellar mass bins $M_{*}=10^{10.6-10.8},10^{10.8-11}\Msun$, a null fraction of unbarred galaxies do not experience a minor merger  against a   fraction of  0.13 and 0.20   in strongly barred galaxies. The large difference is present at times after the bar formed.  The  fraction of unbarred galaxies with at least one minor merger is 0.05, 0.28 and 1 for the stellar mass bins $M_{*} =10^{10.4-10.6},10^{10.6-10.8} ,10^{10.8-11}\Msun$ respectively.  These fractions are systematically higher than the ones of strongly barred galaxies ( 0.04,0.17,0.20 ). 
 
 The results suggest that disk galaxies, whether they have a bar or not, have quiet merger histories and dominated by minor mergers \citep[see also][]{rodriguezgomez2015,izquierdo2019}. This is found in the three  bins of stellar mass. We also find that the frequency of unbarred galaxies that experience at least one minor merger seems to be higher than the one of strongly barred galaxies at times after the bar forms.

\section{Discussion}
\label{sec:discussion}
In the previous sections, we presented the properties of bars found in the $z=0$ disc galaxies of the \TNG\ simulation. We studied their evolution as well as the properties of their host galaxies. In parallel, we analysed the evolution of a control sample of unbarred disc galaxies, with a similar stellar mass and morphology at $z=0$. In this section, we bring together all our results to discuss what we can conclude about the conditions that lead to bar formation and strengthening and about the role of bars in quenching the central regions of the host galaxies.  We conclude the section with an outlook on planned future work. 

\subsection{What are the conditions  to form a strong bar in a disc galaxy?}

In section~\ref{sec:evolgal}, we have seen that during the onset of bar formation ($-0.2<\tnorm<0$) the stellar disc component has already assembled (see Fig.~ \ref{fig:gal_evolution}) with periods of intense star formation. By the time the bar is fully formed, a large disc has already grown and is by far the dominant component of the galaxy, with typical bulge-to-total ratios below $0.2$. The galaxy morphology in the strongly barred sample remains largely unaltered afterwards. The lack of a sufficiently large stabilising bulge component could be one of the key ingredients for the development of strong bars \citep[e.g.,][]{kataria2018}. 
Disc galaxies that do not present a bar at $z=0$ have a much different morphological evolution, with a much less dominant disc component at early times. 
The fact  that bars form in early assembling discs could imply that the location in the cosmic web could be a key factor in setting the physical conditions that favour the development of bar instabilities. \cite{mendez2012} study the bar fraction as a function of stellar mass for three different environments from the field to  Virgo and Coma clusters. They find that the bar fraction in clusters, overall, is higher than that in the field. Also, the peak in the bar fraction as a function of the stellar mass is shifted towards massive galaxies for barred galaxies in clusters than those in the field. \cite{mendez2012} interpret their results as evidence of the possible effects of the environment in the bar formation. 

Together with the early growth of the large scale disc, nuclear feedback can also contribute to limiting the bulge growth and lead to bar growth. 
\cite{bonoli2016} and   \cite{zana2018b} analyse different runs of the Eris suite of simulations and concluded that central feedback from an accreting black hole or star-forming regions could be important in shaping galaxy morphology and setting the dynamical conditions that lead to bar formation. 

In  Fig.~\ref{fig:bhevolution} we show the average time evolution of various properties of the supermassive black holes (SMBHs), their energy rate released kinematically into their surroundings for both barred and unbarred galaxies. 
We can see that the median mass of SMBHs hosted by the strongly barred galaxies (first row) is systematically higher at high redshift than the one hosted by unbarred galaxies. Early growth of the central SMBH, connected to larger AGN feedback at an early time, could contribute in maintaining the bulge of barred galaxies small. 
As we will further discuss, a detailed analysis of feedback processes and the small scale physics happening during the phase of bar formation will be the topic of a follow-up paper that will take advantage of the higher resolution of the TNG50 simulation.

\subsection{ What drives  the quenching of the central part in strongly barred galaxies?}
\label{subsec:quenching}

\begin{figure*}
\includegraphics[width=2\columnwidth]{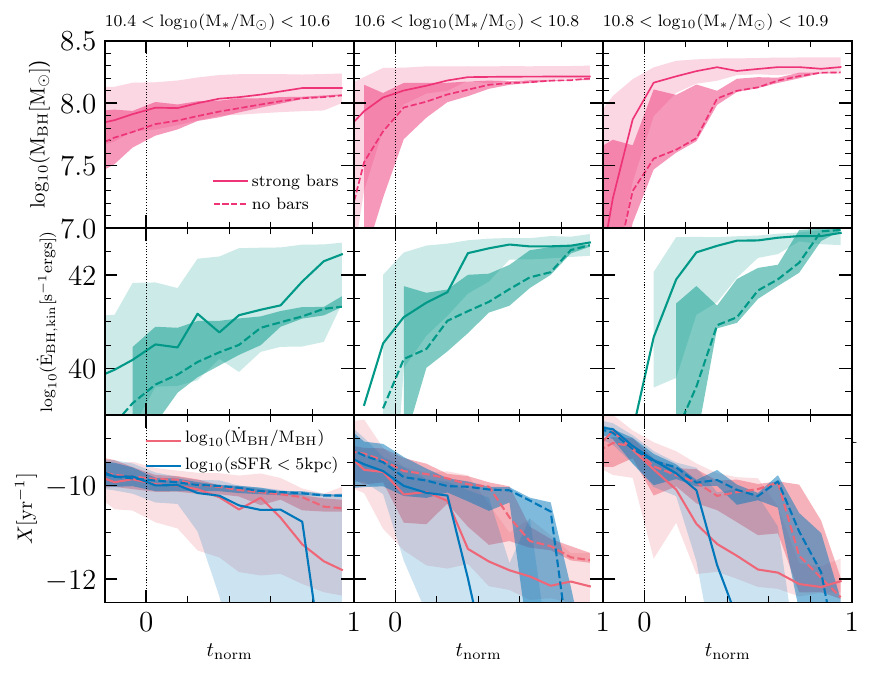}
\caption{The  evolution of SMBHs. From left to right increases with stellar mass. \textit{First row:} The median growth of SMBHs as a function of time hosted by $z=0$ strongly barred (solid lines) and $z=0$ unbarred galaxies (dashed lines) in the TNG100 simulation. \textit{Second row:} The median energy rate kinematically injected by the SMBH  into the surrounding interstellar medium (kinetic mode, green).  \textit{Third row:} The median specific SMBH accretion rate and sSFR within 5 kpc as a function of time. The coloured regions represent the 20$^{\rm th}$ and 80$^{\rm th}$ percentiles of each distribution. The figure shows the possible links between the growth of SMBHs, the formation of the bar, and the subsequent star formation quenching of the nuclear region.}
\label{fig:bhevolution}
\end{figure*}

Once the bar forms and becomes stable, the non-axisymmetric gravitational potential is expected to affect the gas inside it.  In an axisymmetric gravitational potential (such as a stellar disc), the gas will eventually follow it and move in circular orbits. The dynamics of the gas becomes more complex in the presence of a bar (e.g., \citealt{athanassoula1992,maciejewski2002,fragkoudi2016}).  A general picture is that the gas is torqued by the bar and funnelled into the inner Lindblad Resonance or pushed towards the outer Lindblad
Resonance.  The gas then will be shocked and will end up in a  rapid burst of star formation or feeding the supermassive black hole located at the galaxy centre \citep{maciejewski2004,fanali2015}. Depending on the evolution of the strength and length of the bar,  rings or spirals will form close to the inner or outer Lindblad Resonance \citep{athanassoula1992b}.

From Fig.~\ref{fig:sb_evolution},   it is clear that strongly barred galaxies present higher star formation rates while the bars assemble, especially for the galaxies in the range of $\rm log_{10}( M_{*}/\Msun) = 10.6-11$. 
Once the strong bars settle down and continue to grow in strength and length,  the SFR sharply decreases  and is more pronounced in the central part of the galaxies ($<2-5\,\kpc$). The drop in the SFR of strongly barred galaxies is clear when compared with the slowly decaying star formation rates of unbarred galaxies across time, at a fixed $z=0$ stellar mass.  
Similarly, Fig.~\ref{fig:sfe} shows that the normalised efficiency of the gas to be converted into stars  declines sharply after  bar formation within the bar extent. In contrast, the star formation efficiency in an aperture equal to the bar length decreases more steadily with time for the unbarred galaxies.  Also, this behaviour in the star formation efficiency is very different in the region outside the bar since it remains mostly flat across time for both strongly barred and unbarred galaxies.

The results described above suggest that  the bar could play a role in the gas exhaustion and  star formation quenching (especially in the nuclear part) of the $z=0$ strongly barred galaxies, consistent with other theoretical results \citep[e.g.,][]{gavazzi2015,kim2017}. \cite*{spinoso2017}, for example, highlight the role of the bar in leaving the central part of the galaxy quenched in the ErisBH simulation. Furthermore, 
\cite*{khoperskov2018} study systematically isolated gas-rich galaxies  with and without a bar and the response of the gas. They find that, in isolated barred galaxies, the star-formation efficiency decreases rapidly in comparison with isolated unbarred galaxies. They also show that the quenching of the nuclear part of their barred galaxies is due to the bar. Note, however, that their study does not include any AGN feedback prescription.

Indeed, other physical processes could be simultaneously affecting the dynamics of the gas, such as supernovae and  AGN feedback. We speculate that, within  the TNG model, bar evolution and AGN feedback may play in concert at determining the central gas content and star formation activity. Fig.~\ref{fig:bhevolution} (second row) displays the median BH energy rate kinematically released into the ISM (\textit{low accretion rate, kinetic wind mode}), $\dot{E}_{\rm BH, kin}$, and the specific BH accretion rate  $(\dot{M}_{\rm BH}/ M_{\rm BH})$ and the central sSFR ($\leq 5$ kpc; third row).  As we can see in the third row, the sSFR and the specific BH accretion rates follow a similar evolution. Note, however, that the fall in sSFR is sharper in comparison with the fall  in specific BH accretion rates for the barred galaxies after bar formation. 
The median energy rate is higher in strongly barred galaxies than the median in unbarred galaxies at all the times. This is expected because of the early growth of the BHs hosted by strongly barred galaxies as previously discussed. Interestingly, \cite{weinberger2018}, by studying the SMBHs of the total population in the TNG100 simulation, find that  the average $\dot{E}_{\rm BH}$ for the SMBHs hosted by galaxies in a similar stellar mass range ($M_{*}=10^{10.5-11}\Msun$)  at $z=0$ is about $10^{42.5}\ergs$  (see their Fig.~1) which is similar to the rates for the  strongly barred sample. However, while their average population is still star-forming (with the average $\rm SFR\sim 0.3\Msun\, \rm yr^{-1}$),   our strongly barred galaxy sample becomes quiescent.  We speculate that AGN feedback  plays a secondary, or at least ancillary, role in the quenching of strongly barred galaxies. However, a more detailed study is required to assess the causality and the interrelationships among bar formation, AGN feedback, and quenching.

\subsection{Outlook}

The resolution of the TNG100 simulation allows to capture the physics above scales of $\sim 1$ kpc. While this is enough for a general analysis of the formation and evolution of bars within the most massive galaxies, simulations with higher resolution are necessary to properly follow the first stages of bar build-up, as well as the fate of the gas in the very nuclear region of galaxies.
In a future work, we will analyse the evolution of bars in the higher resolution run of the IllustrisTNG suite, TNG50 \citep{pillepich2019,nelson2019}. With this simulation, we can both probe smaller scale dynamics as well as explore the bar population in the lower galaxy mass regimes. By exploiting model variations for different assumptions on AGN feedback along with the fiducial TNG model, we will try to disentangle the physical mechanisms contributing to the formation and evolution of a bar as well as their role in the quenching of the nuclear region. We will try to answer questions as: do bars form because AGN feedback produces the right physical conditions for their formation? Is the presence of the bar stimulating gas accretion onto BHs and hence enhancing their feedback? Would there be quenching via bar formation without BH feedback at all? Namely, does BH feedback really have a secondary role in the quenching of strongly barred galaxies? Or rather is bar formation able to trigger BH feedback also at low BH masses?

Another interesting aspect not included in this work is the analysis of the physical processes that lead to bar destruction. By construction, we have analysed here only bars that survive until the present epoch. However, we expect a large population of bars that develop at an early time and do not survive until today.
The processes advocated for bar weakening or, even destruction, are (i) dynamical interactions with external objects, such as mergers or fly-bys \citep[e.g.,][]{zana2018b,peschken2019} that have been shown to be often, either temporarily, or definitively, destructive for non-axisymmetric structures; (ii) a bar suicide mechanism, in which the strong gas inflows driven by the bar itself are able to deeply reshape the galactic potential and make it, consequently, bar-stable \citep[e.g.,][]{norman1996}; and (iii) bar buckling mechanism that makes the bar evolve on the perpendicular axis, via vertical instabilities \citep[e.g.,][]{combes1981,debattista2004}.
Higher resolution simulations  are needed to properly analyse these kinds of dynamical processes \citep[e.g.,][]{merritt1994,sellwood1994} that will be an important subject of a future analysis.

Finally, we have pointed out how local feedback processes, such as from SNe or AGN, can influence dramatically the physical conditions of the galaxies, thus setting favourable or adverse conditions for bar formation (e.g., \citealt{zana2018c}). The parallel analysis of large cosmological simulations with different feedback prescriptions could offer important insight into how local processes couple with the conditions set by the large scale environment and influence the development of bar structures.

\section{Summary}
\label{sec:summary}
In this paper, we investigate the properties and the buildup of strongly barred galaxies using the $\Lambda$CDM  
magneto-cosmological hydrodynamical simulation \TNG~ from the IllustrisTNG project.  The simulation evolves a comoving region of 110.7 $\rm cMpc$ on a side with  an initial number of particles and gas cells of $2\times1820^3$ and does so with a model for galaxy formation physics which reasonably reproduces  many galaxy observables \citep{pillepich2018a,nelson2018a,naiman2018,springel2018,marinacci2018}. 
The resolution of the simulation allows the exploration of bar formation in a cosmological context, something that was out of reach until now.  Our study concentrates on well-resolved galaxies (with more than $10^4$ stellar particles, or $M_{*}> 10^{10.4} \Msun$ within $\rhalf$).  From the entire galaxy sample, we select $270$ disc galaxies with clear morphology and with a dominant disc component  ($D/T\geq0.5$), as obtained through a kinematic decomposition \citep{genel2015}.  
 
 To identify barred galaxies, we Fourier decompose the face-on stellar surface density and calculate the second term as a function of the cylindrical radius. We define the peak of the amplitude, $\Amax$, as a proxy for the bar strength and the location of the peak as a proxy for the bar length $\rbar$. We consider to be barred galaxies  all disc galaxies with $\Amax\geq0.2$,
 $\rbar>1\,\rm kpc$ and with a phase of $\Ato$  constant inside $\rbar$.  
 These are our main findings on the bar properties and the parent galaxies are:
 \begin{itemize}
 \item   $40$ per cent   of our disc galaxies are barred, with $22$ per cent having strong bars and $18$ per cent weak bars. These bar fractions are in reasonable agreement with observational studies \citep{barazza2008,nair2010} and with previous theoretical studies \citep{algorry2017}. We also find that the bar fraction increases with increasing stellar mass as concluded by several observational studies. (Fig.~\ref{fig:galfracstarmass}).  Massive galaxies tend to have older bars, with $50$ per cent of our strong bars 
  having formed between $z=0.5$ and $z=1.5$.  
 
 \item The bar lengths in our sample span between 2 and $6\,\kpc$ (Fig.~\ref{fig:bardistribution}). We compare the relation between the bar length and galaxy sizes to observational estimates \citep{gadotti2011}, finding a reasonable agreement between them (Fig.~\ref{fig:gassize}). 
 
  \item Barred galaxies exhibit lower gas-to-stellar mass ratios compared to the unbarred sample, and the difference is significantly more pronounced for more massive galaxies (Fig.~\ref{fig:ssfrdiagram}).  Similarly, looking at the sSFR-stellar mass  diagram, barred galaxies with a stellar mass larger than $10^{10.6}\Msun$ have significantly lower sSFRs than their unbarred counterparts, with typical values of quenched galaxies. Unbarred galaxies, instead, tend to be found as star-forming galaxies (Fig.~\ref{fig:ssfrdiagram}).  
 \end{itemize}
 
 To investigate the processes that lead to bar formation and quenching of barred galaxies at $z=0$, 
 we track the history of strongly barred galaxies and compare them to the history of the unbarred sample. 
 Our main results are:
 \begin{itemize}
 \item The majority of $z=0$ strong bars develop between $z=0.5$ and  $z=1$, with their age depending on stellar mass (with a large scatter). After the formation of the bar, its strength and its length grow steadily until reaching the local value  (Fig.~\ref{fig:sb_evolution}). 
 
 \item Before the epoch of bar formation, barred galaxies are active with high star formation rates in the nuclear region ($<2\,\kpc$). Once the bar grows enough, their nuclear star formation rates drop rapidly to becoming \textit{quenched}. Unlike strongly barred galaxies,  the nuclear star formation in unbarred galaxies decreases at a slow rate with time.
  
  \item We find differences in the build-up of strongly barred galaxies compared to the one of unbarred galaxies (Fig.~\ref{fig:gal_evolution}), also at fixed galaxy stellar mass. Strongly barred galaxies rapidly develop a prominent disc and a small spheroidal component that promotes the formation of the bar.  On the contrary, the unbarred galaxies develop a slightly more massive spheroidal component and a smaller disc. It also takes a longer time for  unbarred galaxies to develop their disc component. Both unbarred and strongly barred galaxies selected to have similar stellar mass at $z=0$, end up with similar morphologies at $z=0$ apart from the most massive galaxies. However, strongly barred galaxies grow more rapidly in stellar mass than their unbarred counterparts, namely their stellar mass assembly occurs at earlier times. Furthermore, they contain less  gas than unbarred galaxies after bar formation. 
  
  \item We compare the star formation efficiency of the gas inside the bar extent and outside the bar (Fig.~\ref{fig:sfe}). We normalise the star formation efficiency by the star formation efficiency of the galaxy within $15\, \kpc$, $\SFEdr$. A sharp drop in the star formation efficiency inside the bar region occurs after bar formation. This effect is more pronounced in more massive barred galaxies. By contrast, the star formation efficiency does not present a change with time outside the bar. We also calculate the star formation efficiency in unbarred galaxies in the inner region. Unlike strongly barred galaxies, the star formation efficiency inside this region gradually decreases with time. Our finding could be interpreted as a signature of rapid quenching in the central region of strongly barred galaxies. 

\item Strongly barred  and unbarred disc galaxies present similar merger histories (Fig.~\ref{fig:mergers}) dominated by minor mergers. Unbarred galaxies  present slightly more active recent merger history than strongly barred galaxies.

\end{itemize}

These results demonstrate, first of all, that the TNG model naturally leads to the formation of strong bars and, secondly, that the formation and growth of strongly barred disc galaxies at $z=0$ significantly differ from unbarred disc galaxies. 
The early assembly history of the galaxies seems to play an important role in setting the physical conditions that lead to bar formation.  The properties of the bar structures at $z=0$ and on the star formation and gas content of the simulated galaxies are in broad agreement with observations \citep{masters2012,cervantes2017,gavazzi2015}. This emphasises the potential of the current generation of cosmological hydrodynamical simulations to study the bar population in a cosmological context. Higher resolution runs will  allow a more detailed analysis of the dynamical build up of bars, the  study of gas evolution and star formation within the extent of the bar as well as  the galaxy as a whole. In upcoming papers, we will exploit the TNG50 simulation to extend these findings to lower stellar masses and to disentangle the causal  relationships between bar formation, BH feedback and star-formation quenching. 


\section*{Acknowledgements}
The authors thank Daniele Spinoso for the productive discussions. We also acknowledge  Bernardo Cervantes-Sodi and Victor Debattista for reading the paper and the useful comments.  YRG acknowledges the support of the European Research Council through grant number ERC-StG/716151. SB and YRG acknowledge  \textit{PGC2018-097585-B-C22, MINECO/FEDER, UE} of the Spanish Ministerio de Economia, Industria y Competitividad.   LH thanks to National Science Foundation of China (11721303) and the National Key R\&D Program of China (2016YFA0400702). DIV acknowledge the support from project \textit{AYA2015-66211-C2-2 MINECO/FEDER, UE} of the Spanish Ministerio de Economia, Industria y Competitividad and the grant \textit{Programa Operativo Fondo Social Europeo de Arag\'{o}n 2014-2020. Construyendo Europa desde Arag\'{o}n}. This project has received funding from the European Union Horizon 2020 Research and Innovation Programme under the Marie Sklodowska-Curie grant agreement No 734374. We thank
contributors to SciPy\footnote{http://www.scipy.org}, Matplotlib\footnote{https://matplotlib.org}, and the Python\footnote{http://www.python.org} programming language.








\appendix


\bsp	
\label{lastpage}
\end{document}